\documentclass[12pt,a4paper]{article}
\usepackage{amsmath}
\usepackage{amsfonts}
\usepackage{mathrsfs}
\usepackage{cite}

\renewcommand{\thesection}
  {\arabic{section}.\hspace{-.5em}}
\renewcommand{\thesubsection}
  {\arabic{section}.\arabic{subsection}.\hspace{-.5em}}
\renewcommand{\thesubsubsection}
  {\arabic{section}.\arabic{subsection}.\arabic{subsubsection}.\hspace
                                                               {-.5em}}

\makeatletter
\renewcommand\section{
  \@startsection{section}{3}{\z@}%
  {-3.25ex\@plus -1ex \@minus -.2ex}%
  {1.5ex \@plus .2ex}%
  {\normalfont\normalsize\bfseries\mathversion{bold}}}
\renewcommand\subsection{
  \@startsection{subsection}{3}{\z@}%
  {-3.25ex\@plus -1ex \@minus -.2ex}%
  {1.5ex \@plus .2ex}%
  {\normalfont\normalsize\bfseries\mathversion{bold}}}
\renewcommand\subsubsection{
  \@startsection{subsubsection}{3}{\z@}%
  {-3.25ex\@plus -1ex \@minus -.2ex}%
  {1.5ex \@plus .2ex}%
  {\normalfont\normalsize\itshape}}
\makeatother

\makeatletter \@addtoreset{equation}{section} \makeatother
\renewcommand{\theequation}{\arabic{section}.\arabic{equation}}

\makeatletter
\renewcommand{\appendix}{
\renewcommand{\thesection}{\Alph{section}.\hspace{-.5em}}
\renewcommand{\thesubsection}
  {\Alph{section}.\arabic{subsection}.\hspace{-.5em}}
\renewcommand{\thesubsubsection}
  {\Alph{section}.\arabic{subsection}.\arabic{subsubsection}.\hspace
                                                             {-.5em}}
\@addtoreset{equation}{section}
\renewcommand{\theequation}{\Alph{section}.\arabic{equation}}
\setcounter{section}{0}}
\makeatother

\newcommand{\Eqn}[1]{&\hspace{-0.5em}#1\hspace{-0.5em}&}
\newcommand{\nn}{\nonumber}
\renewcommand{\[}{\begin{equation}}
\renewcommand{\]}{\end{equation}}
\newcommand{\eqb}{\begin{eqnarray}}
\newcommand{\eqe}{\end{eqnarray}}

\newcommand{\varth}{\vartheta}
\newcommand{\hE}{\hat{E}} 
\newcommand{\tE}{\tilde{E}}
\newcommand{\tj}{\tilde{\jmath}}
\newcommand{\tq}{\tilde{q}}
\newcommand{\btau}{{\bar\tau}} 
\newcommand{\ttau}{\tilde{\tau}}
\newcommand{\tDelta}{\tilde{\Delta}}
\newcommand{\calF}{F} 
\newcommand{\hcalF}{{\cal F}} 
\newcommand{\chcalF}{{\check{\cal F}}}
\newcommand{\Gen}{X} 
\newcommand{\dphi}{\partial_\phi}
\newcommand{\bE}{\xi} 
\newcommand{\dbE}{\partial_\bE}
\newcommand{\vecm}{{\boldsymbol{\mu}}} 
\newcommand{\vecw}{{\boldsymbol{w}}}
\newcommand{\veclambda}{{\boldsymbol{\lambda}}}
\newcommand{\veczero}{{\boldsymbol{0}}}
\newcommand{\Hecke}[1]{{\mathscr H}\!\left\{#1\right\}}
\newcommand{\ETheta}{\Theta} 
\newcommand{\bvec}[1]{\boldsymbol{#1}}
\newcommand{\WeylEei}{W(E_8)} 

\newcommand{\tprod}{\prod\nolimits}

\newcommand{\bbC}{{\mathbb C}}
\newcommand{\bbR}{{\mathbb R}}
\newcommand{\bbZ}{{\mathbb Z}}
\newcommand{\bbP}{{\mathbb P}}

\newcommand{\grp}[1]{\mathrm{#1}}

\begin{document}


\def\papertitlepage{\baselineskip 3.5ex \thispagestyle{empty}}
\def\preprinumber#1#2{\hfill
\begin{minipage}{1.2in}
#1 \par\noindent #2 
\end{minipage}}
\renewcommand{\thefootnote}{}
\newcounter{aff}
\renewcommand{\theaff}{\fnsymbol{aff}}
\newcommand{\affiliation}[1]{
  \setcounter{aff}{#1} $\rule{0em}{1.2ex}^\theaff\hspace{-.4em}$}
%
%
\papertitlepage
\setcounter{page}{0}
\preprinumber{YITP-11-94}{arXiv:1111.3967}
\vskip 1ex
~
\vfill
\begin{center}
{\large\bf\mathversion{bold}
Topological string amplitudes for the local $\frac{1}{2}$K3 surface
}
\end{center}
\vskip 1ex
\vfill
\baselineskip=3.5ex
\begin{center}
  Kazuhiro Sakai\\

{\small
\vskip 6ex
  {\it Yukawa Institute for Theoretical Physics, Kyoto University}\\
  {\it Kyoto 606-8502, Japan}\\
\vskip 2ex
{\tt ksakai@yukawa.kyoto-u.ac.jp}

}
\end{center}
\vskip 1ex
\vfill
\baselineskip=3.5ex
\begin{center} {\bf Abstract} \end{center}

We study topological string amplitudes
for the local $\frac{1}{2}$K3 surface.
We develop a method of computing
higher-genus amplitudes
along the lines of the direct integration formalism,
making full use of the Seiberg--Witten curve
expressed in terms of modular forms and 
$E_8$-invariant Jacobi forms.
The Seiberg--Witten curve was constructed previously for 
the low-energy effective theory of
the non-critical E-string theory in $\bbR^4\times T^2$.
We clarify how the amplitudes
are written as polynomials
in a finite number of generators
expressed in terms of the Seiberg--Witten curve.
We determine the coefficients of the polynomials by solving
the holomorphic anomaly equation and the gap condition,
and construct the amplitudes explicitly up to genus three.
The results encompass topological string amplitudes
for all local del Pezzo surfaces.

\vskip 1ex
\vspace*{\fill}
\noindent
November 2011
\setcounter{page}{0}
\newpage
\renewcommand{\thefootnote}{\arabic{footnote}}
\setcounter{footnote}{0}
\setcounter{section}{0}
\baselineskip = 3.5ex
\pagestyle{plain}
%

\section{Introduction}

Topological string theory on the local $\frac{1}{2}$K3 surface
provides us with a unified description of the low-energy
effective theory
of four-dimensional ${\cal N}=2\ \grp{SU}(2)$ gauge theories
\cite{Seiberg:1994aj,Seiberg:1994rs}
and their extensions to five and six dimensions.
The local $\frac{1}{2}$K3 surface is a non-compact
Calabi--Yau threefold
in which the $\frac{1}{2}$K3 surface appears as a divisor.
By blowing down exceptional curves,
one can reduce $\frac{1}{2}$K3
to any del Pezzo surface ${\cal B}_n\ (n\le 8)$,
including $\bbP^2$ and $\bbP^1\times\bbP^1$.
Topological string theory on the local $\frac{1}{2}$K3
describes the low-energy effective theory
of the six-dimensional $(1,0)$ supersymmetric
non-critical E-string theory
in $\bbR^4\times T^2$
\cite{Ganor:1996mu,Seiberg:1996vs,Klemm:1996hh,Ganor:1996xd,
      Ganor:1996pc,Minahan:1998vr}.
Similarly, topological string theory on the local ${\cal B}_n$
corresponds to the non-critical $E_n$ string theory
in $\bbR^4\times T^2$
with one of the cycles of the $T^2$ shrinking to zero size
\cite{Lerche:1996ni,Minahan:1997ch}.
This theory shares the same moduli space with
the five-dimensional ${\cal N}=1\ \grp{SU}(2)$ gauge theory
on $\bbR^4\times S^1$
with $n-1$ fundamental matters
\cite{Seiberg:1996bd,Morrison:1996xf,Douglas:1996xp}.
For the toric case $(n\le 5)$,
the topological string amplitudes have been well studied.
In particular, the all-genus topological string partition function
in this case is given by the Nekrasov partition function
for the above five-dimensional gauge theory
\cite{Nekrasov:2002qd,Nekrasov:2003rj,Iqbal:2003zz,Eguchi:2003sj}.

For toric Calabi--Yau threefolds,
the construction of topological string amplitudes
has been well understood.
One can use the topological vertex formalism
\cite{Aganagic:2003db}
to construct the all-genus partition function
as a sum over partitions
on the A-model side.
The ``remodeling the B-model'' conjecture
\cite{Bouchard:2007ys},
which is based on the topological recursion
for matrix models
\cite{Eynard:2007kz},
enables us to generate the amplitudes on the B-model side
recursively with respect to the genus
\cite{Eynard:2010dh,Eynard:2010vd}.
Indeed, for toric local del Pezzo surfaces,
topological string amplitudes have been
studied both in the former approach
\cite{Konishi:2003qq}
and in the latter approach
\cite{Brini:2008rh,Bouchard:2008gu,Manabe:2009sf}.
For non-toric Calabi--Yau threefolds,
however, such a universal prescription is lacking
at present.
The purpose of this paper is to
formulate a method of constructing
the topological string amplitudes
for the most general local $\frac{1}{2}$K3 surface.

A generalization of the topological vertex formalism was proposed
\cite{Diaconescu:2005ik}
and applied to the construction of the
topological string partition functions
for non-toric local del Pezzo surfaces
\cite{Diaconescu:2005mv}.
(See also \cite{Konishi:2006ya} for another construction
for the local ${\cal B}_6$.)
Remarkably, this formalism enables us to construct
the all-genus partition function
as a sum over partitions.
The partition function in this form is, however, not suitable
for obtaining
the topological string amplitude at each genus
in a closed form.
Also these constructions do not seem to apply directly to
the case of the general local $\frac{1}{2}$K3 surface.
On the other hand, 
one can construct the topological string amplitude
at each genus by solving
the holomorphic anomaly equation \cite{Bershadsky:1993cx}.
Higher-genus amplitudes have been constructed explicitly
for some special cases with one or two moduli parameters
\cite{Katz:1999xq,Hosono:1999qc,Mohri:2001zz}.
Moreover,
a simple, specific form
of the holomorphic anomaly equation
was proposed
for the topological string amplitudes
for the local $\frac{1}{2}$K3 surface
\cite{Minahan:1998vr,Hosono:1999qc}.
By solving this equation one can construct
higher-genus amplitudes for the most general case
with manifest affine $E_8$ symmetry
\cite{Minahan:1998vr,Hosono:2002xj}.
In this construction, however, the amplitudes are obtained
not in a closed form, but rather in the form of
an instanton expansion with respect to one of
the K\"ahler moduli parameters.

Recently, it has been discovered and proved that
topological string amplitudes
for any Calabi--Yau threefold
are polynomials in a finite number of generators
\cite{Yamaguchi:2004bt,Alim:2007qj}.
By making use of this remarkable fact
and taking account of the symmetry,
in particular modular properties of the amplitudes
\cite{Aganagic:2006wq},
one can directly solve the holomorphic anomaly equation
and efficiently determine the amplitude at each genus in a closed form
\cite{Grimm:2007tm}.
This method, which we will call the direct integration method,
is applicable,
in principle, to topological strings
on any Calabi--Yau threefold.
It has also been applied to the
gravitational corrections to
Seiberg--Witten theories
\cite{Huang:2006si,Grimm:2007tm,Huang:2009md,Huang:2011qx}.

There are many examples of non-compact Calabi--Yau threefolds
for which the mirror geometries are essentially
described by Seiberg--Witten curves.
In this case, the symmetry of the topological string
amplitudes can naturally be understood
in terms of the Seiberg--Witten curve.
The Seiberg--Witten curve turns out to be useful
to construct the topological string amplitude
not only at genus zero, but also at higher genus.
All these arguments apply to
the local $\frac{1}{2}$K3 surface:
The mirror geometry in this case
is described by the Seiberg--Witten curve for the E-string theory
\cite{Ganor:1996pc,Eguchi:2002fc}.
In particular, the most general form expressed in terms of
modular forms and $E_8$-invariant Jacobi forms
was constructed \cite{Eguchi:2002fc}.
Making full use of this Seiberg--Witten curve,
we are able to formulate a method of constructing
the topological string amplitudes
at higher genus in a closed form
for the most general local $\frac{1}{2}$K3.

Let us briefly summarize our construction
in the following.
We first clarify the polynomial structure of the higher-genus
amplitudes and identify the generators of the polynomials.
The generators are expressed in terms of
one of the periods and the complex structure modulus
of the torus associated with the Seiberg--Witten curve.
We elucidate the modular anomaly of the generators,
which can be interpreted as the holomorphic anomaly.
This enables us to evaluate the holomorphic anomaly
of the ans\"atze for the higher-genus amplitudes.
Each time we solve the holomorphic anomaly equation,
there appears a holomorphic ambiguity that cannot be fixed
by the equation. We fix them by imposing a gap condition.
The gap condition for the topological strings on
the local $\frac{1}{2}$K3 surface is known
\cite{Minahan:1998vr}.
This comes from the geometric property of the local $\frac{1}{2}$K3.
Using this method,
we construct the amplitudes
explicitly up to genus three.

While the basic idea of our construction
is the same as that of the direct integration method,
ours is rather different from the standard one
in appearance.
We start from the holomorphic anomaly equation
of Hosono--Saito--Takahashi
\cite{Hosono:1999qc} specific to the present model,
rather than that of Bershadsky--Cecotti--Ooguri--Vafa (BCOV)
\cite{Bershadsky:1993cx}.
We use our original generators
when constructing ans\"atze for the amplitudes.
In terms of these generators
the amplitudes can be concisely expressed.
Despite these differences,
both methods should be essentially equivalent.
We show that the amplitudes
and the holomorphic anomaly equation
can be written in a form akin to what have been obtained
for other models
by the standard direct integration method
\cite{Aganagic:2006wq,Huang:2006si,Grimm:2007tm,Huang:2009md}.

As we mentioned in the beginning, the topological string theory
on the local $\frac{1}{2}$K3 surface encompasses
that on all local del Pezzo surfaces.
Remarkably, when the topological string amplitudes
for the local $\frac{1}{2}$K3 are
expressed in terms of the Seiberg--Witten curve,
their forms are universal to all local del Pezzo surfaces.
To obtain the amplitudes for any local del Pezzo surface,
we have only to reduce the Seiberg--Witten curve correspondingly
\cite{Minahan:1997ch,Eguchi:2002nx}.
By way of illustration,
we present explicit forms of amplitudes
for three basic examples:
the massless local ${\cal B}_8$, the local $\bbP^2$,
and the local $\bbP^1\times\bbP^1$.

This paper is organized as follows.
In section 2, we review some basic properties
of the topological string amplitudes
for the local $\frac{1}{2}$K3 surface.
In section 3, we describe the method of constructing
topological string amplitudes for the local $\frac{1}{2}$K3
in a closed form. First we review 
how the topological string amplitude at genus zero
is constructed from the Seiberg--Witten curve.
We then study the modular anomaly of fundamental quantities
and interpret them as the holomorphic anomaly.
With these data, we solve the holomorphic anomaly equation
at low genus. We make a conjecture on the general
structure of the amplitudes, which greatly simplifies
the problem of solving the holomorphic anomaly equation.
We present two other expressions for the amplitudes
and the holomorphic anomaly equation. In particular,
the last expression is similar to what is found in
the standard direct integration method.
In section 4, we study how to reduce our general results
to the topological string amplitudes
for all local del Pezzo surfaces.
We present explicit forms of amplitudes
for three basic examples:
the massless local ${\cal B}_8$, the local $\bbP^2$,
and the local $\bbP^1\times\bbP^1$.
Section 5 is devoted to the conclusion and discussion.
In Appendix A, we present explicitly
the generators of $E_8$-invariant Jacobi forms
and the Seiberg--Witten curve for the present model.
Appendix B is a collection of derivative formulas.
In Appendix C,
we present the explicit form of the amplitude at genus three.
In Appendix D, we summarize our conventions of special functions.

\section{Properties of topological string amplitudes
         for local $\frac{1}{2}$K3}

In this section we review some basic
properties of the topological string amplitudes for the local
$\frac{1}{2}$K3 surface.
The reader is referred to references
\cite{Minahan:1998vr,Hosono:1999qc,Mohri:2001zz}
for further details.

The $\frac{1}{2}$K3 surface, also known as
the rational elliptic surface
or the almost del Pezzo surface ${\cal B}_9$,
is obtained by blowing up nine base points
of a pencil of cubic curves in $\bbP^2$.
The $\frac{1}{2}$K3 surface admits
an elliptic fibration over $\bbP^1$.
A generic $\frac{1}{2}$K3 surface has 12 singular fibers,
while a generic elliptic K3 surface has 24 singular fibers.

The second homology group $H_2(\frac{1}{2}{\rm K3},\bbZ)$
is generated by the class of a line in $\bbP^2$
and the nine classes of the exceptional curves.
With an inner product given by the intersection number,
$H_2(\frac{1}{2}{\rm K3},\bbZ)$ acquires
the structure of
the ten-dimensional
odd unimodular Lorentzian
lattice $\Gamma^{9,1}$
(also denoted by ${\rm I}_{9,1}$).
The automorphism group of $\Gamma^{9,1}$ contains
the Weyl group of the affine $E_8$ root system.
This property is crucial to our construction of
the topological string amplitudes
for the local $\frac{1}{2}$K3.
It is also useful to note that the lattice decomposes as
$\Gamma^{9,1}=\Gamma^{1,1} \oplus \Gamma_8$,
where $\Gamma^{1,1}$ is
the two-dimensional
odd unimodular Lorentzian lattice
and $\Gamma_8$ is the $E_8$ root lattice.
$\Gamma^{1,1}$ is generated by
$[\rm B],[\rm E]$ with
$[\rm B]\cdot[\rm E]=1,\ [\rm B]\cdot[\rm B]=-1,\ 
 [\rm E]\cdot[\rm E]=0$,
where $[\rm B]$ and $[\rm E]$ can be viewed as
the classes of the base and the fiber
of the elliptic fibration.
The automorphism group of $\Gamma_8$ is given by
the Weyl group of the $E_8$ root system,
which will be denoted by $\WeylEei$.

By a local $\frac{1}{2}$K3 surface we mean
the total space of the canonical bundle
of a generic $\frac{1}{2}$K3 surface.
It is a non-compact Calabi--Yau threefold.
We consider the A-model topological string theory on it.
In this paper we let $\calF_g$ denote
{\it the instanton part} of
the topological string amplitude at genus $g$.
What we mean by
{\it the instanton part} will be explained soon.
We consider the amplitudes in real polarization,
namely, $\calF_g$ are holomorphic functions.
As we will see below, the holomorphic anomaly of the amplitudes
can be read from the modular anomaly.

Let $\calF$ denote the all-genus
topological string partition function defined as
\[
\calF=\sum_{g=0}^\infty \calF_g x^{2g-2}.
\]
$\calF$ can be viewed as the generating function of
the Gopakumar--Vafa invariants \cite{Gopakumar:1998jq}.
By taking account of
the $\WeylEei$ symmetry,
$F$ can be expressed as
\eqb
\calF(\varphi,\tau,\vecm;x)
\Eqn{=}
 \sum_{r=0}^\infty\sum_{n=1}^\infty\sum_{k=0}^\infty
 \sum_{\veclambda\in P_+}\sum_{\vecw\in {\cal O}_\veclambda}
 \!\!N^r_{n,k,\veclambda}
 \sum_{m=1}^\infty\frac{1}{m}
 \left(2\sin\frac{mx}{2}\right)^{2r-2}
 e^{2\pi i m(n\varphi+k\tau+\vecw\cdot\vecm)}.\nn\\
\eqe
Here, $P_+$ denotes the set of all dominant weights of $E_8$,
and the sum with respect to weights $\vecw$ is taken
over the Weyl orbit of $\veclambda$;
$\varphi$ and $\tau$ denote the K\"ahler moduli corresponding to 
the base and the fiber of the elliptic fibration, respectively,
while $\vecm=(\mu_1,\ldots,\mu_8)$ denote
the orthogonal
coordinates for the complexified root space of $E_8$.
The Gopakumar--Vafa invariants $N^r_{n,k,\veclambda}$
are integers.
They count the BPS multiplicities of
the five-dimensional ${\cal N}=1$ supersymmetric theory
obtained by compactifying the M-theory
on the local $\frac{1}{2}$K3 surface.
This five-dimensional theory is identified with
the effective theory of
the six-dimensional E-string theory
on $\bbR^5\times S^1$.

We defined $\calF_g$ as {\it the instanton part},
which means that $\calF_g$ is expanded as
\[\label{FginZ}
\calF_g(\varphi,\tau,\vecm)
 = \sum_{n=1}^\infty Z_{g,n}(\tau,\vecm)\, e^{2\pi i n \varphi}
\]
and does not contain
any polynomial (including constant) term in $\varphi$.
From the point of view of the E-string theory,
$Z_n := e^{-\pi i n \tau}Z_{0,n}$
is the BPS partition function of the E-strings wound $n$ times
\cite{Klemm:1996hh, Minahan:1998vr}. $Z_n$ is also interpreted as
the partition function of ${\cal N}=4\ \grp{U}(n)$
topological Yang--Mills theory on
$\frac{1}{2}$K3 \cite{Minahan:1998vr}.
Throughout this paper, we refer to this $\calF_g$
as the topological string amplitude at genus $g$.

At present, the most general way of
computing higher-genus amplitudes
applicable to any Calabi--Yau threefold
is to solve the BCOV holomorphic anomaly equation
\cite{Bershadsky:1993cx}.
In this paper we define the topological string amplitudes
as holomorphic functions,
but one could adopt the standard definition
in terms of twisted ${\cal N}=2$ superconformal field theories,
in which the amplitudes also possess anti-holomorphic dependence
on moduli parameters.
It is well known that this anti-holomorphic dependence,
or the holomorphic anomaly,
is governed by
the BCOV holomorphic anomaly equation
\[\label{BCOV}
{\bar{\partial}}_{\bar{\imath}}
 F_g
=\frac{1}{2}
\bar{C}_{\bar{\imath}\bar{\jmath}\bar{k}}e^{2K}
G^{j\bar{\jmath}}G^{k\bar{k}}\left(
D_jD_kF_{g-1}+\sum_{h=1}^{g-1}D_jF_hD_kF_{g-h}
\right).
\]
Here, $F_g$ denotes the amplitudes at genus $g$,
$K$ is the K\"ahler potential,
$G_{i\bar{\jmath}}=\partial_i\bar{\partial}_{\bar{\jmath}}K$
is the K\"ahler metric,
$D_i$ denotes a certain covariant derivative,
and $\bar{C}_{\bar{\imath}\bar{\jmath}\bar{k}}=\overline{C_{ijk}}$
with $C_{ijk}=D_iD_jD_kF_0$.
One can recursively solve this differential equation
to construct higher-genus amplitudes $F_g$ up to holomorphic
ambiguities.
It is worth noting that $F_g$ are polynomials
in a finite number of generators
\cite{Yamaguchi:2004bt,Alim:2007qj},
which greatly helps the construction.

In practice, however, it is rather hard to solve
a topological string model with ten K\"ahler moduli parameters,
in particular when the target space is not
a local toric Calabi--Yau threefold.
Nevertheless,
in the case of the local $\tfrac{1}{2}$K3
one can make full use of the symmetry
to construct the amplitudes much more efficiently
than in generic cases.
As is explained below,
$\calF_g$ at low $g$ are fully characterized by
the symmetry,
the holomorphic anomaly equation
and the gap condition.

Let us start with the symmetry.
Due to the automorphism of the homology lattice of $\frac{1}{2}$K3,
the partition function exhibits
the affine $E_8$ symmetry.
Moreover, it possesses good modular properties in $\tau$.
It is known that $Z_{g,n}$ has
the following structure \cite{Mohri:2001zz}:
\[\label{ZinT}
Z_{g,n}(\tau,\vecm)
=\frac{T_{g,n}(\tau,\vecm)}
      {\left[\prod_{k=1}^\infty(1-q^k)\right]^{12n}},
\]
where
\[
q=e^{2\pi i\tau}.
\]
$T_{g,n}$ is a $\WeylEei$-invariant quasi-Jacobi form
of weight $2g-2+6n$ and index $n$.
The reader is referred to Appendix A for the basic properties of the
$\WeylEei$-invariant Jacobi form.
By  $\WeylEei$-invariant {\it quasi}-Jacobi forms
we mean those which are generated
by the generators of the ordinary
$\WeylEei$-invariant Jacobi forms
and the Eisenstein series $E_2(\tau)$.

$E_2(\tau)$ is not strictly a modular form,
as it transforms as
\[\label{E2Stransf}
E_2\left(-\frac{1}{\tau}\right)
  =\tau^2\left(E_2(\tau)+\frac{6}{\pi i\tau}\right).
\]
However, the non-holomorphic function
\[
\hE_2(\tau,\btau):=E_2(\tau)+\frac{6}{\pi i(\tau-\btau)}
\]
transforms as a modular form of weight 2,
\[
\hE_2\left(-\frac{1}{\tau},-\frac{1}{\btau}\right)
  =\tau^2\hE_2(\tau,\btau).
\]
By replacing all $E_2(\tau)$ by $\hE_2(\tau,\btau)$,
the amplitude $\calF_g$ transforms as a modular function
of weight $2g-2$
at the cost of losing holomorphicity.
This non-holomorphicity
is regarded as the holomorphic anomaly
of the amplitude.
In other words,
the modular/holomorphic anomaly of the amplitude
always appears through $E_2$.
For later convenience,
we introduce a normalized notation
$\bE := \frac{1}{24}E_2$
and let
\[
\dbE=24\partial_{E_2}
\]
measure the holomorphic anomaly.
We also introduce a normalized variable
$\phi = 2\pi i\varphi + \phi_0$, so that
\[
\dphi = \frac{1}{2\pi i}\partial_\varphi.
\]
The precise relation between $\phi$ and $\varphi$
will be given in section 3.
Throughout this paper we hold $\tau$ and $\vecm$
constant
when we take partial derivatives with respect to $\bE$ and $\phi$.
In terms of these normalized variables,
the holomorphic anomaly equation
for the partition function $F$ is written as
\cite{Hosono:1999qc}
\[
\dbE e^\calF
  =x^2\dphi(\dphi+1)e^\calF.
\]
By expanding the equation in $x$,
it becomes a set of recursive equations:
\eqb
\label{HAEforFg}
\dbE\calF_g
 \Eqn{=}\dphi^2\calF_{g-1}+\dphi\calF_{g-1}
  +\sum_{h=0}^{g}\dphi\calF_h\dphi\calF_{g-h}.
\eqe
The equation for $g=0$ should be understood
with $\calF_{-1}=0$.
In terms of $Z_{g,n}$, the holomorphic anomaly equations read
\[
\label{HAEforZgn}
\dbE Z_{g,n}
=n(n+1)Z_{g-1,n}
 +\sum_{h=0}^{g}\sum_{k=1}^{n-1} k(n-k)Z_{h,k}Z_{g-h,n-k}.
\]
Again, the equation for $g=0$ should be understood
with $Z_{-1,n}=0$.

The above form of holomorphic anomaly equation was first proposed
for $g=0$
\cite{Minahan:1998vr}
and later extended
for general $g$ \cite{Hosono:1999qc}.
The validity of the equation has been further confirmed in
\cite{Mohri:2001zz, Hosono:2002xj}.
It is expected that
the above equation is equivalent to
the BCOV holomorphic anomaly equation
for the local $\tfrac{1}{2}$K3 \cite{Hosono:2002xj, Hosono:2008np}.

As the holomorphic anomaly equation is a differential equation,
one needs to fix the integration constant,
i.e.~the holomorphic ambiguity, at each genus.
For the present model, it is known that the following gap condition
can be used for this purpose:
\[
\calF
 =\sum_{n=1}^\infty e^{2\pi in\varphi}\left(
  \frac{1}{n(2\sin\frac{nx}{2})^2}+{\cal O}(q^n)\right).
\]
This condition is equivalent to
the following constraint on the Gopakumar--Vafa invariants:
\eqb
N^g_{n,k,\veclambda}\Eqn{=}0\quad\mbox{for}\quad k<n
\quad\mbox{except}\quad N^0_{1,0,\veczero}=1.
\eqe
This follows from
the geometric structure
of the local $\frac{1}{2}$K3 \cite{Minahan:1998vr}.
In terms of $Z_{g,n}$ the gap condition reads
\[\label{gapcondZ}
Z_{g,n}=\beta_g n^{2g-3}+{\cal O}(q^n),
\]
where $\beta_g$ are rational numbers defined
by the following expansion:
\eqb\label{beta_g}
\sum_{g=0}^\infty\beta_g x^{2g}
\Eqn{=}\frac{x^2}{4\sin^2\frac{x}{2}}\nn\\
\Eqn{=}1+\frac{1}{12}x^2+\frac{1}{240}x^4+\frac{1}{6048}x^6
 +{\cal O}(x^8).
\eqe

It has been checked \cite{Minahan:1998vr, Mohri:2001zz, Hosono:2002xj}
for low $g$ and $n$ that
$Z_{g,n}$ can be determined uniquely
by the symmetry (\ref{ZinT}),
the holomorphic anomaly equations (\ref{HAEforZgn}),
and the gap conditions (\ref{gapcondZ}).\footnote{
For general $g$, however, these conditions are not likely
to be sufficient for determining the amplitude completely.
See the discussion at the end of subsection~3.3.}
Based on this fact,
we will develop a method of constructing $\calF_g$
in a closed form in the next section.

\section{Closed expressions for amplitudes}

\subsection{Genus zero amplitude and instanton expansion}

It is known that the genus zero amplitude $\calF_0$ for
the local $\frac{1}{2}$K3 surface
is obtained as the prepotential
associated with the
Seiberg--Witten curve of the form
\[\label{geneE8curve}
y^2=4x^3-fx-g,
\]
with
\[\label{genefg}
f=\sum_{j=0}^4 a_j u^{4-j},\qquad
g=\sum_{j=0}^6 b_j u^{6-j}.
\]
Actually, a Seiberg--Witten curve of this form
itself describes an elliptic fibration of the $\frac{1}{2}$K3 surface.
It can be viewed as a sort of local mirror symmetry
between one $\frac{1}{2}$K3 and
another $\frac{1}{2}$K3 \cite{Minahan:1998vr, Mohri:2001zz}.
We present the explicit form of the Seiberg--Witten curve
in Appendix A.
It was determined in \cite{Eguchi:2002fc}
so that the instanton expansion of the prepotential
correctly reproduces
$Z_{0,n}$ at low $n$ calculated by the method
of \cite{Minahan:1998vr},
which we summarized in the last section.

Let us recall how the prepotential is obtained from
the Seiberg--Witten curve
of the above general form.
Given the Seiberg--Witten curve (\ref{geneE8curve}),
the expectation value of the scalar component of the ${\cal N}=2$
vector multiplet is expressed as
\[\label{SWvev}
\phi=-\frac{1}{2\pi}\int du\oint_{\alpha}\frac{dx}{y},
\]
where $\alpha$ is one of the fundamental cycles of the curve.
The complexified gauge coupling constant
$\ttau$ is given by the complex structure modulus of
the Seiberg--Witten curve. On the other hand,
$\ttau$ is given by the second derivative of
the prepotential. In terms of the instanton part $\calF_0$
of the prepotential,
$\ttau$ is expressed as
\[\label{SWttau}
\ttau=\tau+\frac{i}{2\pi}\partial_\phi^2\calF_0,
\]
where $\tau$ is the bare gauge coupling constant.
By solving these relations, one obtains the prepotential
from the Seiberg--Witten curve.

The practical calculation can be organized
as follows \cite{Minahan:1997ch,Minahan:1997ct,Eguchi:2002fc}.
Since the present Seiberg--Witten curve is elliptic,
one can make full use of the explicit map between an elliptic curve
and a torus.
Let $(2\pi\omega,2\pi\omega\ttau)$
denote the fundamental periods of the torus.
The map from the torus to the elliptic curve in the Weierstrass form
(\ref{geneE8curve}) is given
in terms of the Weierstrass $\wp$-function by
\[
x=\wp(z;2\pi\omega,2\pi\omega\ttau),\qquad
y=\partial_z\wp(z;2\pi\omega,2\pi\omega\ttau).
\]
The coefficients of the elliptic curve
and the periods of the torus
are related as
\[
\label{fg_tauomega}
f=\frac{1}{12}\frac{\tE_4}{\omega^4},\qquad
g=\frac{1}{216}\frac{\tE_6}{\omega^6},
\]
where we use the notation
\[
\tE_{2n}:=E_{2n}(\ttau).
\]

One can express $\ttau$ and $\omega$
in terms of the Seiberg--Witten curve
by inverting the modular functions.
First, we eliminate $\omega$
from the two equations (\ref{fg_tauomega})
by taking the ratio $f^3/g^2$.
Equivalently, we can look at the $j$-invariant.
We expand it in $u^{-1}$ as
\eqb
\frac{1}{\tj}
\Eqn{=}\frac{\tE_4^3-\tE_6^2}{1728\tE_4^3}
 =\frac{f^3-27g^2}{1728f^3}\nn\\
\Eqn{=}\frac{1}{j}
 -\frac{E_6 b_1}{4E_4^3}
  \frac{1}{u}+{\cal O}\left(\frac{1}{u^2}\right).
\label{tj_tq}
\eqe
Here we have used
$a_0=\frac{1}{12}E_4,\ b_0=\frac{1}{216}E_6,\ a_1=0$.\footnote{
This is the convention of the Seiberg--Witten curve
adopted in \cite{Eguchi:2002fc}.
By suitable rescaling and shift of variables
one can always recast a generic Seiberg--Witten curve of the form
(\ref{geneE8curve}), (\ref{genefg}) as this form
without loss of generality.}
On the other hand, the $j$-invariant has the following expansion:
\[
\tj = \frac{1}{\tq} +744 +196884\tq 
+{\cal O}\left(\tq^2\right),\qquad \tq=e^{2\pi i\ttau}.
\]
Inverting this expansion and using (\ref{tj_tq}), we obtain
the expansion of $\ttau$ in $u^{-1}$.
By introducing the notation
\[\label{t_def}
t:=2\pi i(\ttau-\tau),
\]
the expansion is expressed as
\[
t=-\frac{E_4b_1}{4\Delta}\frac{1}{u}
 +\left(
   \frac{E_6 a_2}{48\Delta}
  -\frac{E_4 b_2}{4\Delta}
  -\frac{E_4(E_4E_2+5E_6){b_1}^2}{192\Delta^2}
   \right)\frac{1}{u^2}
 +{\cal O}\left(\frac{1}{u^3}\right).
\label{t_u}
\]
Substituting this into (\ref{fg_tauomega}),
one obtains
the expansion of $\omega$ in $u^{-1}$.
We choose the sign of $\omega$
in such a way that $\omega$ is expanded as
\[\label{omegainu}
\omega
=\frac{1}{u}
 -\frac{(E_4 E_2-E_6)b_1}{48\Delta}\frac{1}{u^2}
 +{\cal O}\left(\frac{1}{u^3}\right).
\]
Integrating this by $u$, one obtains $\phi$.
We define $\phi$ with the normalization
\[
\label{phidef}
\phi:=-\int\omega du
\]
so that $e^\phi$ has the expansion
\[\label{expphi_u}
e^\phi = \frac{1}{u}
 -\frac{(E_4 E_2-E_6)b_1}{48\Delta}\frac{1}{u^2}
 +{\cal O}\left(\frac{1}{u^3}\right).
\]
Inverting this relation, we have
\[\label{mirrormap}
\frac{1}{u}=e^\phi
 +\frac{(E_4 E_2-E_6)b_1}{48\Delta}e^{2\phi}
 +{\cal O}\left(e^{3\phi}\right).
\]
Substituting this into (\ref{t_u}), we obtain
\[\label{tinphi}
t=-\frac{E_4b_1}{4\Delta}e^\phi
 +\left(\frac{E_6a_2}{48\Delta}-\frac{E_4b_2}{4\Delta}
  -\frac{E_4(E_4E_2+2E_6)b_1^2}{96\Delta^2}
  \right)e^{2\phi}
 +{\cal O}\left(e^{3\phi}\right).
\]
Similarly, from (\ref{omegainu}) and (\ref{mirrormap})
we obtain
\eqb
\label{loinphi}
\ln\omega
\Eqn{=}\phi
 +\left(\frac{(E_6E_2-E_4^2)a_2}{1152\Delta}
  -\frac{(E_4E_2-E_6)b_2}{96\Delta}\right.\nn\\
&&\hspace{3em}\left.
{}+\frac{(-2E_4^2E_2^2-8E_6E_4E_2+5E_4^3+5E_6^2)b_1^2}{9216\Delta^2}
  \right)e^{2\phi}
 +{\cal O}\left(e^{3\phi}\right),\qquad
\eqe
which will be used later.
As explained in the beginning, the instanton part of the prepotential
is given by
\[
\label{F0sol}
\calF_0 = -\dphi^{-2}t.
\]
Here $\dphi^{-1}$ denotes
integration with respect to $\phi$
of a power series in $e^\phi$.

This prepotential
is identified with the genus zero amplitude $\calF_0$
for the local $\frac{1}{2}$K3 surface.
The bare gauge coupling $\tau$ is interpreted as
the K\"ahler modulus $\tau$
of the original $\frac{1}{2}$K3.
The scalar expectation value $\phi$ is identified with
the K\"ahler modulus $\varphi$
as \cite{Minahan:1997ct}
\[\label{phivarphirel}
e^\phi
= -q\left[\tprod_{k=1}^\infty(1-q^k)\right]^{12} e^{2\pi i\varphi}.
\]
Taking this into account, $\calF_0$ is expanded as
\eqb
F_0
\Eqn{=}
-\frac{E_4 b_1}{4\left[\prod_{k=1}^\infty (1-q^k)\right]^{12}}
 e^{2\pi i\varphi}\nn\\
&&+\frac{-2E_6\Delta a_2+24E_4\Delta b_2+(E_4^2E_2+2E_6E_4)b_1^2}
  {384\left[\prod_{k=1}^\infty (1-q^k)\right]^{24}}
 e^{4\pi i\varphi}
 +{\cal O}\left(e^{6\pi i\varphi}\right).
\eqe
By substituting the coefficients $a_n,b_n$
of the Seiberg--Witten curve presented in Appendix A,
one obtains the genus zero amplitude for the local $\frac{1}{2}$K3
as a series expansion in $e^{2\pi i\varphi}$
up to any desired order.

\subsection{Modular anomaly}

The Seiberg--Witten curve transforms as
a $\WeylEei$-invariant Jacobi form (see Appendix A).
On the other hand, the genus zero amplitude $\calF_0$
contains $E_2$ and therefore exhibits the modular anomaly.
The $E_2$'s appear when one expands the $j$-invariant $j(\ttau)$
around $\ttau=\tau$. Thus,
the modulus $\ttau$ and the period $\omega$
of the Seiberg--Witten curve do exhibit
the modular anomaly when expanded in $u^{-1}$.
In \cite{Minahan:1997ct},
the modular anomaly of $\ttau$ and $\omega$
was studied
in the course of proving the holomorphic anomaly equation
for the genus zero amplitude.
Extending the analysis,
here we study the modular anomaly
of various quantities derived from the Seiberg--Witten curve.
We will use this to solve
the holomorphic anomaly equation for higher-genus amplitudes.

As mentioned above, the Seiberg--Witten curve
transforms as a Jacobi form.
This means that the modulus $\ttau(u,\tau,\vecm)$
of the curve transforms in precisely the same way
as $\tau$ does under the action of $\grp{SL}(2,\bbZ)$.
It then follows that $t=2\pi i(\ttau-\tau)$
is invariant under $\tau\to\tau+1,\ \ttau\to\ttau+1$,
while it transforms as
\[
\frac{1}{t}
  \to\tau^2\left(\frac{1}{t}+\frac{1}{2\pi i\tau}\right)
\qquad\mbox{for}\qquad
\tau\to -\frac{1}{\tau},\ \ttau\to -\frac{1}{\ttau}.
\]
This anomalous behavior is expected since $E_2$'s appear
in the coefficients of the expansion (\ref{t_u}).
Moreover, one finds that the transformation of $t^{-1}$
is very similar to that of $E_2$ as in (\ref{E2Stransf}).
This suggests that $t^{-1}$ depends on $E_2$ as
\[
t^{-1}=\frac{1}{12}E_2+\mbox{(modular function of weight $2$)}.
\]
One can explicitly check this using the series expansion (\ref{t_u}).
Let us express it as
\[
\left(\dbE t^{-1}\right)_u = 2,
\]
or
\[\label{HAoft}
\left(\dbE t\right)_u = -2t^2.
\]
Here, $\left(\dbE t\right)_u$ denotes
the partial derivative of $t$ with respect to $\bE$,
holding $u$ constant.

Next let us consider modular properties of the combination
\[
\omega t = \left(\frac{\tE_4}{12f}\right)^{1/4}t.
\]
We have used (\ref{fg_tauomega}). 
One can see that this
transforms as a modular form of weight 4 in $\tau$,
since the constituents transform
as
\[
\tE_4\to\ttau^4\tE_4,\qquad
f\to\tau^{-20}f,\qquad
t\to\ttau^{-1}\tau^{-1}t
\]
under the S-transformation $\tau\to -1/\tau,\ \ttau\to -1/\ttau$.
This means that the combination
$\omega t$ is free of the modular anomaly, namely
\[
\left(\dbE(\omega t)\right)_u=0.
\]
Using (\ref{HAoft}), one obtains
\[\label{HAofomega}
\left(\dbE\omega\right)_u=2\omega t.
\]
Furthermore, combining (\ref{HAofomega}) with
(\ref{phidef}) one obtains
\eqb
\label{HAofphi}
\left(\dbE\phi\right)_u\Eqn{=}2\dphi^{-1}t,\\
\label{HAofphi2}
\left(\dbE^2\phi\right)_u\Eqn{=}0.
\eqe
Based on these formulas and (\ref{fg_tauomega}),
one can evaluate the modular anomaly of various quantities.
We present a list of formulas in Appendix B.

So far in this subsection,
we have regarded $u$ and $\bE$ as independent variables
and taken the derivative $\dbE$ holding $u$ constant.
Let us say we are in the $(u,\bE)$ frame.
On the other hand, the holomorphic anomaly equations (\ref{HAEforFg})
are given in the $(\phi,\bE)$ frame.
It is useful to see how expressions
in these frames are transformed into each other.
The derivative of a function $A$ with respect to $\bE$
is transformed between these frames by the simple chain rule
\eqb
\left(\dbE A\right)_\phi\Eqn{=}
  -(\dbE \phi)_u\,\left(\dphi A\right)_{\bE}
  +\left(\dbE A\right)_u,\\
\Eqn{=}\label{E2chainrule}
  -2(\dphi^{-1}t)\,\left(\dphi A\right)
  +\left(\dbE A\right)_u.
\eqe
We have used (\ref{HAofphi}) in the second equality.
We sometimes omit the subscript $\bE$, as we always
hold $\bE$ constant when we take
derivatives $\partial_u$ and $\dphi$.
Applying this formula to $\dphi\calF_0=-\dphi^{-1}t$
and using (\ref{HAofphi}), (\ref{HAofphi2}),
we see that
\[
\left(\dbE\left(\dphi\calF_0\right)\right)_\phi
=2\left(\dphi\calF_0\right)\left(\dphi^2\calF_0\right).
\]
By integrating both sides by $\phi$,
we obtain
the holomorphic anomaly equation (\ref{HAEforFg})
at $g=0$
\[
\dbE\calF_0=\left(\dphi\calF_0\right)^2.
\]
%

\subsection{Higher-genus amplitudes}

The expression (\ref{HAEforFg})
of the holomorphic anomaly equation
is not convenient for practical purposes,
since derivatives of $\calF_g$ appear on both sides of the equation.
Using $\dphi\calF_0=-\dphi^{-1}t$
and the chain rule (\ref{E2chainrule}),
one can rewrite the equation into the recursive form
\[\label{HAEhybrid}
\left(\dbE\calF_g\right)_u
= \dphi^2\calF_{g-1}+\dphi\calF_{g-1}
  +\sum_{h=1}^{g-1}\dphi\calF_h\dphi\calF_{g-h}
\]
for $g\ge 1$.
In the following, we solve this equation
and construct $\calF_g$ for low $g$.

Let us first consider the case of $g=1$.
In this case, the equation simply reads
\eqb
\left(\dbE\calF_1\right)_u
\Eqn{=}\dphi^2\calF_0+\dphi\calF_0\nn\\
\Eqn{=}-t-\dphi^{-1}t.
\eqe
With the help of the derivative formulas
(\ref{diffE2first})--(\ref{diffE2last}),
one immediately finds a solution of the form
\eqb
\calF_1\Eqn{=}c_1\ln\omega
  -\left(\frac{c_1}{12}+\frac{1}{24}\right)\ln\tDelta
  -\frac{1}{2}\phi+f_1(\tau).
\eqe
The constant $c_1$ and the function $f_1(\tau)$
can be determined
by the condition that
$\calF_1$ takes the form (\ref{FginZ}),
namely it does not contain any polynomial term in $\phi$.
From (\ref{tinphi}) and (\ref{loinphi}) we see that
\[
\ln\tDelta=\ln\Delta+{\cal O}(e^{2\pi i\varphi}),\qquad
\ln\omega=\phi+{\cal O}(e^{4\pi i\varphi}).
\]
Using these 
we can determine the unknowns as
$c_1=1/2,f_1=(\ln\Delta)/12$ and obtain
\eqb\label{F1sol}
\calF_1\Eqn{=}\frac{1}{2}\ln\omega
  -\frac{1}{12}\ln\tDelta+\frac{1}{12}\ln\Delta-\frac{1}{2}\phi.
\eqe
While this is not a rigorous derivation,
we have checked that the above form is the correct answer.
Combined with (\ref{mirrormap})--(\ref{loinphi})
and (\ref{phivarphirel}),
the above expression correctly reproduces $Z_{1,n}$,
which we explicitly calculated up to $n=5$ using the method
explained in the last section.
It also reproduces the result for $\vecm=\veczero$
presented in \cite{Hosono:2002xj}.
Note that a similar expression has been presented
for four-dimensional Seiberg--Witten theories
\cite{Eguchi:2003sj, Huang:2009md}.

To compute amplitudes for $g\ge 2$ by solving (\ref{HAEhybrid}),
we point out an interesting
fact that the term $-\phi/2$ in $\calF_1$
precisely cancels the linear term
$\dphi\calF_{g-1}$ on the right-hand side of (\ref{HAEhybrid}).
Therefore, if we introduce the notation
\[\label{FcalFrel}
\hcalF_1 = \calF_1+\frac{1}{2}\phi,\qquad
\hcalF_2 = \calF_2+\frac{1}{96}E_2,\qquad
\hcalF_g = \calF_g\quad\mbox{for}\quad g\ge 3,
\]
the holomorphic anomaly equation (\ref{HAEhybrid})
turns into the very simple form
\[\label{HAEhomogen}
\left(\dbE\hcalF_g\right)_u
= \dphi^2\hcalF_{g-1}
  +\sum_{h=1}^{g-1}\dphi\hcalF_h\dphi\hcalF_{g-h}
\]
for $g\ge 2$.
Note that this form
has already been presented in \cite{Huang:2009md}
in the case of four-dimensional
$\grp{SU}(2)$ Seiberg--Witten theories.
It is natural that the holomorphic anomaly equation
takes the same form in the present case,
since the definition (\ref{phidef}) of
$\phi$ through the Seiberg--Witten curve is
common in both cases.

Based on this simple form,
let us construct the amplitude at $g=2$.
Equation (\ref{HAEhomogen}) in this case reads
\eqb
(\dbE\hcalF_2)_u
\Eqn{=}
   \dphi^2\hcalF_1
  +(\dphi\hcalF_1)^2\nn\\
\Eqn{=}
  \frac{1}{2}\dphi^2\ln\omega+\frac{1}{4}(\dphi\ln\omega)^2
  -\frac{1}{12}\tE_2\,\dphi t\,\dphi\ln\omega
  -\frac{1}{12}\tE_2\,\dphi^2 t
  +\frac{1}{144}\tE_4(\dphi t)^2.\nn\\
\label{HAEforF2}
\eqe
Note that the last expression is a polynomial in
(quasi-)modular forms $\tE_{2k}$
and derivatives $\dphi^m\ln\omega,\,\dphi^n t$.
The polynomial is constrained so that
each term contains two $\dphi$'s
and is of weight 0.
Note that after every $E_2$ is replaced by $\hE_2$,
a modular function in $\ttau$ transforms
as that in $\tau$ with the same weight.
Thus,
the weights of the generators of the polynomial
read
\[
[\tE_{2k}]=2k,\qquad [\dphi^m\ln\omega]=0,\qquad [\dphi^n t]=-2.
\]
We see from (\ref{HAEforF2}) that
$\hcalF_2$ is of weight 2, since $\bE$ is of weight 2.
Let us make an ansatz that
$\hcalF_2$ has the same polynomial structure
as (\ref{HAEforF2}),
namely a polynomial in $\tE_{2k},\,\dphi^m\ln\omega,\,\dphi^n t$
with two $\dphi$'s. 
Explicitly, the ansatz reads
\eqb
\hcalF_2\Eqn{=}
  c_1\tE_2\,\dphi^2\ln\omega
 +c_2\tE_2(\dphi\ln\omega)^2
 +(c_3\tE_2^2+c_4\tE_4)\dphi t\,\dphi\ln\omega\nn\\
&&
 +(c_5\tE_2^2+c_6\tE_4)\dphi^2 t
 +(c_7\tE_2^3+c_8\tE_4\tE_2+c_9\tE_6)(\dphi t)^2.\qquad
\eqe
Substituting this ansatz into (\ref{HAEforF2}),
one can partly determine the coefficients $c_j$.
The derivatives of the generators
with respect to $\bE$
are summarized in Appendix~B.
One has to be careful when taking derivatives
of $\dphi^n\ln\omega$ and $\dphi^n t$ with respect to $\bE$.
We differentiate them in the $(u,\bE)$ frame,
where $\dbE$ and $\dphi$ do not commute.
The explicit forms of these derivatives for general $n$
are given in (\ref{diffE2bisfirst}), (\ref{diffE2bislast}),
which can be shown by using the chain rule (\ref{E2chainrule}).

The holomorphic anomaly equation (\ref{HAEforF2})
reduces the number of undetermined parameters to three.
These remaining parameters can be fixed
by the condition that $\calF_2$ takes the form (\ref{FginZ})
and by the gap conditions (\ref{gapcondZ}) at $n=1,2$.
In the end, one obtains
\eqb\label{F2sol}
\hcalF_2\Eqn{=}
  \frac{1}{48}\tE_2\,\dphi^2\ln\omega
 +\frac{1}{96}\tE_2(\dphi\ln\omega)^2
 -\frac{1}{576}(\tE_2^2-\tE_4)\dphi t\,\dphi\ln\omega\nn\\
&&
 -\frac{1}{1920}(5\tE_2^2+3\tE_4)\dphi^2 t
 -\frac{1}{207360}(35\tE_2^3+51\tE_4\tE_2-86\tE_6)(\dphi t)^2.\qquad
\eqe

In the same way,
we are able to determine the amplitude at genus three.
The most general ansatz for $\hcalF_3$ is written with
68 unknown parameters.
The holomorphic anomaly equation
gives 45 relations and
leaves 23 undetermined parameters.
These are fixed completely by the condition that
$\calF_3$ takes the form (\ref{FginZ})
and by the gap conditions (\ref{gapcondZ}) up to $n=4$.
The explicit form of $\calF_3$ is presented in Appendix~C.
We checked for low $n$ that $Z_{g,n}$ calculated from
the above-obtained $\hcalF_2,\hcalF_3$
are in agreement with the results obtained
by the method described in section 2.
In the next section 
we will also reproduce the higher-genus amplitudes for
local del Pezzo surfaces from these results,
which serves as another consistency check.

Based on the above explicit construction of $\hcalF_g$
at low genera, we propose the following conjecture:
\[
\label{Fgstructureinot}
\fbox{
\begin{minipage}{.85\textwidth}
$\hcalF_g\ (g\ge 2)$ is a polynomial in
$\tE_{2k},\,\dphi^m\ln\omega,\,\dphi^n t
 \,\ (k=1,2,3,\ m,n\in\bbZ_{>0})$,
in which each term
contains $2g-2$ $\dphi$'s
and
is of weight $2g-2$.
\end{minipage}
}
\]
Note that the form of the polynomial
is no longer unique for $g\ge 4$,
since not all of $\dphi^m\ln\omega,\ \dphi^n t$
are independent.
Actually, they are finitely generated.
This can be seen as follows.
Recall that $f,g$ are polynomials of
degree $4,6$ in $u$, respectively.
Since $\partial_u=-\omega\dphi$, this means that
\eqb
\label{WTIdE4}
\left(\omega\dphi\right)^k
\frac{\tE_4}{\omega^4}\Eqn{=}0,\qquad\mbox{for}\quad k>4,\\
\label{WTIdE6}
\left(\omega\dphi\right)^k
\frac{\tE_6}{\omega^6}\Eqn{=}0,\qquad\mbox{for}\quad k>6.
\eqe
These relations give rise to non-trivial relations among
the derivatives $\dphi^m\ln\omega,\ \dphi^n t$.
Using these relations, one can express
all $\dphi^m\ln\omega$ and $\dphi^n t$
with $m,n\in\bbZ_{>0}$ in terms of those
with $m=1,\dots,6$ and $n=1,\ldots,4$
and $\tE_2,\tE_4,\tE_6$.
Therefore, by assuming the above conjecture,
$\hcalF_g$ can also be expressed in terms of
these generators.
In this expression $\hcalF_g\ (g\ge 4)$
is still a polynomial in
$\dphi^m\ln\omega$ and $\dphi^n t$,
but becomes a rational function in $\tE_{2k}$.
In subsection 3.5 we will introduce another expression
in which $\hcalF_g$ is indeed written as a polynomial
of a finite number of generators.

The polynomial structure of $\hcalF_g$ is expected,
since topological string amplitudes
for any Calabi--Yau threefold
are polynomials in a finite number of generators \cite{Alim:2007qj}.
The significance of the conjecture
(\ref{Fgstructureinot}) is that
the generators are explicitly given
in terms of the Seiberg--Witten curve.
This allows us to study topological string amplitudes
for not only the local $\tfrac{1}{2}$K3 but
all local del Pezzo surfaces in a unified way,
as we will see in the next section.
The conjecture (\ref{Fgstructureinot})
provides us with a systematic construction
of the ansatz for general $\hcalF_g$.
In particular, the holomorphic anomaly equations
and the gap conditions will be sufficient
for completely fixing the form of $\hcalF_g$
at low genus, as we have explicitly seen for $g=2,3$.

On the other hand, it is not likely that
these conditions suffice to determine
$\hcalF_g$ at general $g$.
In fact, if we apply the method to
the special case with $\vecm=\veczero$,
where $T_{g,n}(\tau,\veczero)$ are
now ordinary quasi-modular forms,
an undetermined coefficient appears already at $g=4$.
To determine the amplitude completely at general $g$,
one needs additional conditions,
such as gap conditions imposed at the conifold loci
of the moduli space.
This is indeed the case for
some simple local del Pezzo surfaces
\cite{Haghighat:2008gw,Alim:2008kp}.

\subsection{Expression in $(u,\bE)$ frame}

Equation (\ref{HAEhomogen}) looks somewhat irregular,
as the left-hand side is written in the $(u,\bE)$ frame
while the right-hand side is in the $(\phi,\bE)$ frame.
For practical purposes, this is actually a convenient form
since the derivatives of $\hcalF_g$
are assembled in a single term in the $(u,\bE)$ frame
while the gap condition can be explicitly expressed
in the $(\phi,\bE)$ frame.
On the other hand, it is also useful to
express the equation entirely in the $(u,\bE)$ frame.
By using $\dphi=-\frac{1}{\omega}\partial_u$,
(\ref{HAEhomogen}) can be rewritten as
\eqb\label{HAEforFginu}
\left(\dbE\hcalF_g\right)_u
\Eqn{=}
  \frac{1}{\omega^2}\left(
   \partial_u^2\hcalF_{g-1}
  -\partial_u\ln\omega\,\partial_u\hcalF_{g-1}
  +\sum_{h=1}^{g-1}\partial_u\hcalF_h\partial_u\hcalF_{g-h}\right)
\quad
\eqe
for $g\ge 2$.
In this frame, 
it is easier to take
derivatives with respect to $\bE$,
while the gap condition cannot be expressed in a simple manner.
$\hcalF_2$ is expressed as
\eqb
\hcalF_2\Eqn{=}\frac{1}{\omega^2}\left(
  \frac{1}{48}\tE_2\,\partial_u^2\ln\omega
 -\frac{1}{96}\tE_2(\partial_u\ln\omega)^2
 +\frac{1}{5760}(5\tE_2^2+19\tE_4)
    \partial_u t\,\partial_u\ln\omega\right.\nn\\
&&\left.\hspace{2em}{}
 -\frac{1}{1920}(5\tE_2^2+3\tE_4)\partial_u^2 t
 -\frac{1}{207360}(35\tE_2^3+51\tE_4\tE_2-86\tE_6)
                  (\partial_u t)^2\right),\nn\\
\eqe
which is almost as simple as the previous expression (\ref{F2sol}).

\subsection{Expression in direct integration style}

There is another interesting expression
of the topological string amplitudes and the holomorphic anomaly
equation. One can express the amplitudes directly in terms of
the coefficients of the Seiberg--Witten curve.
To see this,
let us start with studying the transformation rules
for the generators.

From (\ref{fg_tauomega}) and 
\[
D:=f^3-27g^2=\frac{\tDelta}{\omega^{12}},
\]
we see that
\eqb
\partial_u\ln f\Eqn{=}\partial_u\ln\tE_4-4\partial_u\ln\omega\nn\\
 \Eqn{=}\frac{1}{3}\left(\tE_2-\frac{\tE_6}{\tE_4}\right)\partial_u t
  -4\partial_u\ln\omega,\\
\partial_u\ln D\Eqn{=}\partial_u\ln\tDelta-12\partial_u\ln\omega\nn\\
 \Eqn{=}\tE_2\,\partial_u t -12\partial_u\ln\omega.
\eqe
Solving these relations, one obtains
\eqb
\label{udifft}
\partial_u t\Eqn{=}
 \frac{1}{\omega^2}\frac{-6fg'+9f'g}{2D},\\
\label{udiffomega}
\partial_u\ln\omega\Eqn{=}
 \frac{(-2fg'+3f'g)\Gen+(-2f^2f'+36gg')}{8D},
\eqe
where
\[
\Gen:=\frac{\tE_2}{\omega^2}.
\]
The derivative of $\Gen$ in $u$ is computed as
\[
\label{udiffX}
\partial_u\Gen=
  \frac{(2fg'-3f'g)\Gen^2+(4f^2f'-72gg')X+(24f^2g'-36ff'g)}{8D}.
\]
With the help of these relations and
$\dphi=-\frac{1}{\omega}\partial_u$,
it is straightforward to express
higher derivatives of $t$ and $\ln\omega$
in terms of $\Gen$
and $f^{(m)}(u),\ g^{(n)}(u)$.
Note that $\tE_4,\tE_6$ can also be rewritten
in terms of $f,\,g,\,\omega$ by using (\ref{fg_tauomega}).
After all this, $\hcalF_2$ is expressed as
\eqb\label{F2infg}
\hcalF_2\Eqn{=}
\frac{1}{92160 D^2}
\left(
(100{f}^2{g'}^2-300{f}{f'}{g}{g'}+225{f'}^2{g}^2)\Gen^3\right.\nn\\
&&\hspace{1em}
+( 240{f}^4{g''}-780{f}^3{f'}{g'}
  -360{f}^3{f''}{g}+990{f}^2{f'}^2{g}\nn\\
&&\hspace{2.15em}
  -6480{f}{g}^2{g''}+11880{f}{g}{g'}^2
  -14580{f'}{g}^2{g'}+9720{f''}{g}^3)\Gen^2\nn\\
&&\hspace{1em}
+(-480{f}^5{f''}+420{f}^4{f'}^2+8640{f}^3{g}{g''}+10080{f}^3{g'}^2
  -54000{f}^2{f'}{g}{g'}\nn\\
&&\hspace{2.15em}
  +12960{f}^2{f''}{g}^2
  +29160{f}{f'}^2{g}^2
  -233280{g}^3{g''}+213840{g}^2{g'}^2)\Gen\nn\\
&&\hspace{1em}
+( 5184{f}^5{g''}
-12816{f}^4{f'}{g'}-7776{f}^4{f''}{g}+15336{f}^3{f'}^2{g}
-139968{f}^2{g}^2{g''}\nn\\
&&\hspace{2.05em}\left.
+258336{f}^2{g}{g'}^2-428976{f}{f'}{g}^2{g'}
+209952{f}{f''}{g}^3+167184{f'}^2{g}^3
)
\right).\hspace{3em}
\eqe
Note that $\omega$ does not appear explicitly in this expression.
The same type of expression for $\calF_3$ is immediately obtained
by rewriting the result in Appendix C.
We do not present its lengthy expression here,
but the calculation is straightforward.

The holomorphic anomaly equations can be written
in a form more suited to the above expression.
Observe that when the $\hcalF_g$ are expressed as in (\ref{F2infg}),
holomorphic anomalies appear only through $\Gen$.
Hence, one can simply replace $\dbE$ by
\[
\dbE=(\dbE\Gen)_u\partial_\Gen=\frac{24}{\omega^2}\partial_\Gen
\]
in the holomorphic anomaly equation (\ref{HAEforFginu}).
For $g=2$, the equation is now written as
\[\label{HAEDI2}
24\partial_\Gen\hcalF_2
=\partial_u^2\hcalF_{1}
 -\partial_u\ln\omega\,\partial_u\hcalF_1
 +\left(\partial_u\hcalF_1\right)^2.
\]
By using
\eqb
\partial_u\hcalF_1
 \Eqn{=}-\frac{1}{2}\partial_u\ln\omega
        -\frac{1}{12}\partial_u\ln D\nn\\
 \Eqn{=}\frac{(2fg'-3f'g)\Gen+(-2f^2f'+36gg')}{16D}
\eqe
and (\ref{udiffomega}), (\ref{udiffX}),
one can evaluate the right-hand side of (\ref{HAEDI2})
as a quadratic polynomial in $\Gen$.
Then, integrating directly both sides of (\ref{HAEDI2}) in $\Gen$,
one obtains (\ref{F2infg}) up to the ``constant'' part in $\Gen$.
For $g\ge 3$, let us introduce the notation
\[
\chcalF_1=\hcalF_1-\frac{1}{2}\ln\omega,\qquad
\chcalF_g=\hcalF_g\quad\mbox{for}\quad g\ge 2.
\]
The holomorphic anomaly equations can then be written
again in a very simple form:
\[\label{HAEDIg}
24\partial_\Gen\chcalF_g
=\partial_u^2\chcalF_{g-1}
 +\sum_{h=1}^{g-1}\partial_u\chcalF_h\partial_u\chcalF_{g-h},
\qquad g\ge 3.
\]

Regarding
the explicit form of $\hcalF_g$ at $g=2,3$ and the above equation,
we present our conjecture on the structure of the amplitudes
in another form:
$\hcalF_g\ (g\ge 2)$ can be expressed as
\[\label{Fgstructureinfg}
\hcalF_g = \frac{1}{D^{2g-2}}\sum_{k=0}^{3g-3}
  P_{g,k}[\partial_u^{2g-2},f,g]\Gen^k.
\]
Here, $P_{g,k}[\partial_u^{2g-2},f,g]$
denotes a polynomial in
$\partial_u^m f,\,\partial_u^n g\ (m,n\in\bbZ_{\ge 0})$
in which each term contains $2g-2$ $\partial_u$'s.
The form of the polynomial is constrained
so that $\hcalF_g$ transforms as
a $\WeylEei$-invariant quasi-Jacobi form of index 0.
Note that the constituents of the amplitudes
are of the following indices (see Appendix A):
\[
[\Gen]=2,\quad [f]=4,\quad [g]=6,\quad [D]=12,\quad [\partial_u]=-1.
\]
Recall that $f,g$ are polynomials of degree $4,6$ in $u$,
respectively.
Therefore, in this expression it is manifest that
$\hcalF_g$ is a polynomial in a finite number of generators,
namely,
\[
\frac{1}{D},\quad \Gen,\quad
\partial_u^m f,\quad
\partial_u^n g,
\qquad
m=0,\ldots,4,\quad n=0,\ldots,6.
\]

We find that
the above structure
of the amplitudes and the holomorphic anomaly equation
is akin to
what has been obtained for other models
by the direct integration method
\cite{Aganagic:2006wq,Huang:2006si,Grimm:2007tm,Huang:2009md}.
While we have taken a different path from the standard approach,
both constructions should be essentially equivalent.

\section{Topological string amplitudes for local del Pezzo surfaces}

The topological string amplitudes
for the local $\frac{1}{2}$K3 surface encompass
those for all local del Pezzo surfaces.
In this section we see
how the former reduce to the latter.
In fact, when 
the topological string amplitudes
for the local $\frac{1}{2}$K3
are expressed in terms of the Seiberg--Witten curve,
their forms
are universal to all local del Pezzo surfaces.
We obtain the amplitudes for any local del Pezzo surface
by merely replacing the Seiberg--Witten curve
with the corresponding one.
The mirror pair of the local del Pezzo surface ${\cal B}_n$
is given by the Seiberg--Witten curve for
the five-dimensional $E_n$ strings \cite{Minahan:1997ch}.
It is also easy to reduce the most general Seiberg--Witten curve
to that for any del Pezzo surface
\cite{Minahan:1997ch,Yamada:1999xr,Eguchi:2002nx}.
We first discuss the general cases
and then
present explicit forms of amplitudes
for three basic examples:
the massless local ${\cal B}_8$, the local $\bbP^2$,
and the local $\bbP^1\times\bbP^1$.

\subsection{General cases}

The Seiberg--Witten curve for the local ${\cal B}_8$
is obtained from that for the local $\frac{1}{2}$K3
by simply taking the limit $q\to 0$.
Curves for the other local ${\cal B}_n\ (n\le 7)$
are immediately obtained by a suitable rescaling
\cite{Minahan:1997ch,Eguchi:2002nx}.
The construction of the topological string amplitudes
from the Seiberg--Witten curve is
essentially the same as in the case of
the local $\frac{1}{2}$K3.
In particular, 
the mirror map between $u$ and $\phi$
for ${\cal B}_n\ (n\le 8)$ is simply given by
the $q\to 0$ limit of (\ref{mirrormap}).

Below we present the minor modifications needed
for the local ${\cal B}_n\ (n\le 8)$.
The instanton parts of
the topological string amplitudes at $g=0,1$
are slightly modified as follows:
\eqb
\label{dPF0}
\calF_0\Eqn{=}-\dphi^{-2}t + \frac{9-n}{6}\phi^3,\\
\label{dPF1}
\calF_1\Eqn{=}\frac{1}{2}\ln\omega-\frac{1}{12}\ln\tDelta
  -\frac{1}{2}\phi
  +\frac{9-n}{12}\phi,
\eqe
with 
\[
t:=2\pi i\ttau
\]
instead of (\ref{t_def}).
Expressions for
higher-genus amplitudes $\hcalF_g\ (g\ge 2)$
hold as they stand, where the $\hcalF_g$ are now
related to $\calF_g$ as
\[
\hcalF_1 = \calF_1+\frac{1}{2}\phi,\qquad
\hcalF_2 = \calF_2+\frac{1}{96},\qquad
\hcalF_g = \calF_g\quad\mbox{for}\quad g\ge 3.
\]
We also need to modify the relation (\ref{phivarphirel})
between $\phi$ and $\varphi$, since it is no longer valid
in the limit $q=0$.
Instead of (\ref{phivarphirel}), we
identify them by
\[
e^\phi
= - e^{2\pi i\varphi}.
\]
%

\subsection{Massless local ${\cal B}_8$}

As an illustration we first consider the case of
local ${\cal B}_8$ with $\vecm=\veczero$.
In this case the corresponding Seiberg--Witten curve
is extremely simple.
The coefficients are given by
\[
f=\frac{1}{12}u^4,\qquad g=\frac{1}{216}u^6-4u^5.
\]
The amplitudes at $g=0,1$ are given by
(\ref{dPF0}), (\ref{dPF1}) with $n=8$.
By substituting the above $f,\,g$ into (\ref{F2infg}) one obtains
\eqb
\hcalF_2\Eqn{=}
 \frac{1}{207360u^4(u-432)^2}
 \Bigl(25\Gen^3+15u(-25u+6048)\Gen^2\nn\\
&&\hspace{2em}+75u^2(29u^2-22464u+5225472)\Gen
+u^5(335u-273888)\Bigr).
\eqe
Similarly, from the expression of $\calF_3$ in Appendix C,
one obtains
\eqb
\hcalF_3\Eqn{=}
 \frac{1}{5016453120u^8(u-432)^4}
 \Bigl(525\Gen^6-8400u^2\Gen^5\nn\\
&&\hspace{2em}
 +315u^2(175u^2+5184u+5225472)\Gen^4\nn\\
&&\hspace{2em}
 +560u^3(-325u^3+18360u^2-89859456u+11851370496)\Gen^3\nn\\
&&\hspace{2em}
 +63u^4(4625u^4-5008896u^3+8491143168u^2\nn\\
&&\hspace{5em}
  -2300402073600u+260052929740800)\Gen^2\nn\\
&&\hspace{2em}
 +672u^7(-325u^3+284796u^2-623837376u+7054387200)\Gen\nn\\
&&\hspace{2em}
 +u^8(61775u^4-96755904u^3+219325750272u^2\nn\\
&&\hspace{5em}
     +15910182715392u+9788763779629056)
\Bigr).
\eqe
From these expressions one can compute Gopakumar--Vafa invariants.
The instanton expansions in this case read
\eqb
\frac{1}{u}\Eqn{=}
 e^{\phi}-60e^{2\phi}-1530e^{3\phi}-274160e^{4\phi}
 -50519055e^{5\phi}+{\cal O}\left(e^{6\phi}\right),\\
\omega\Eqn{=}
 e^{\phi}+5130e^{3\phi}+1347520e^{4\phi}+372046365e^{5\phi}
 +{\cal O}\left(e^{6\phi}\right),\\
t\Eqn{=}\phi
 +252e^{\phi}+36882e^{2\phi}+7637736e^{3\phi}+1828258569e^{4\phi}
+{\cal O}\left(e^{5\phi}\right).
\eqe
The Gopakumar--Vafa invariants are computed
by recasting $F$ as
\[
\sum_{g=0}^\infty \calF_g x^{2g-2}
=
 \sum_{r=0}^\infty\sum_{n=1}^\infty
 N^r_{n}
 \sum_{m=1}^\infty\frac{1}{m}
 \left(2\sin\frac{mx}{2}\right)^{2r-2}
 e^{2\pi i mn\varphi}.
\]
\begin{table}[t]
\[
\begin{array}{|c|crrrrrr|}
\hline
N^r_n&n&1&2&3&4&5&\cdots\\ \hline
r&&&&&&&\\[1ex]
0&& 252 & - 9252 & 848628 & - 114265008 &  18958064400 &\\
1&& -2 & 760 &  - 246790 &  76413833 &  - 23436186176 &\\
2&& 0 & -4 & 30464 &  - 26631112 &  16150498760 &\\
3&& 0 & 0 & -1548 &  5889840 &  - 7785768630 &\\
\vdots&&&&&&&\ddots\\ \hline
\end{array}\nn
\]
\caption{\label{table:B8GVinv}Gopakumar--Vafa
invariants for the massless local ${\cal B}_8$.}
\end{table}

We present the Gopakumar--Vafa invariants $N^r_{n}$ at low degrees
in Table \ref{table:B8GVinv}.
This reproduces the known result,
for example found in \cite{Klemm:1996hh,Mohri:2001zz}.\footnote{
The Gopakumar--Vafa invariants $N^r_n$ at $r=1$
and the instanton numbers $\tilde{N}^g_n$
for $g=1$ curves found in \cite{Klemm:1996hh,Mohri:2001zz}
are related by $N^1_n = \sum_{k|n}\tilde{N}^1_{(n/k)}$
\cite{Mohri:2001zz}.}
Moreover, it is easy to compute $N^r_{n}$ up to an arbitrarily large
degree of $n$,
as we now have the exact form of the amplitudes $\hcalF_g$.

We have performed the expansion around
the large volume point $u=\infty$ to compute the Gopakumar--Vafa
invariants, but we could expand the amplitudes at
arbitrary $u$.
It would be interesting to study the behavior of
the amplitudes around the other points such as
the orbifold point, as in \cite{Bouchard:2008gu}.

\subsection{Local $\bbP^2$}

The coefficients of the Seiberg--Witten curve are given by
\[
f=\frac{1}{12}u^4-2u,\qquad g=\frac{1}{216}u^6-\frac{1}{6}u^3+1.
\]
The amplitudes at $g=0,1$ are given by
(\ref{dPF0}), (\ref{dPF1}) with $n=0$.
By substituting the above $f,\,g$ into (\ref{F2infg}), one obtains
\eqb
\hcalF_2\Eqn{=}
 \frac{75\Gen^3-165u^2\Gen^2+125u^4\Gen+9(5u^6-464u^3+6192)}
      {7680(u^3-27)^2}.
\eqe
Similarly, from the expression of $\calF_3$ in Appendix C,
one obtains
\eqb
\hcalF_3\Eqn{=}
 \frac{1}{20643840(u^3-27)^4}
 \Bigl(14175\Gen^6-75600u^2\Gen^5
 +315u(533u^3+3024)\Gen^4\nn\\
&&\hspace{2em}
 -560(355u^6+6750u^3+8748)\Gen^3\nn\\
&&\hspace{2em}
 +21u^2(6305u^6+257472u^3+1181952)\Gen^2\nn\\
&&\hspace{2em}
 -672u(70u^9+5007u^6+49086u^3+34992)\Gen\nn\\
&&\hspace{2em}
 +6965u^{12}+774992u^9+13201920u^6+27993600u^3+20155392
\Bigr).\qquad
\eqe

The instanton expansions in this case are given by
\eqb
\frac{1}{u}\Eqn{=}
 e^{\phi} - 2 e^{4\phi} - e^{7\phi} - 20 e^{10\phi}
 - 177 e^{13\phi} 
 +{\cal O}\left(e^{16\phi}\right),\\
\omega\Eqn{=}
 e^{\phi} + 4 e^{4\phi} + 41 e^{7\phi} + 520 e^{10\phi}
 + 7275 e^{13\phi} 
 +{\cal O}\left(e^{16\phi}\right),\\
t\Eqn{=}9\phi
 +27 e^{3\phi} + \frac{405}{2} e^{6\phi} + 2196 e^{9\phi}
 + \frac{110997}{4} e^{12\phi} 
 +{\cal O}\left(e^{15\phi}\right).
\eqe
The all-genus topological string partition function
can be expressed as
\[
\sum_{g=0}^\infty \calF_g x^{2g-2}
=
 \sum_{r=0}^\infty\sum_{n=1}^\infty
 N^r_{n}
 \sum_{m=1}^\infty\frac{1}{m}
 \left(2\sin\frac{mx}{2}\right)^{2r-2}
 Q^{mn},
\]
where
\[
Q = e^{6\pi i\varphi} = -e^{3\phi}.
\]
Table \ref{table:P2GVinv} shows
the Gopakumar--Vafa invariants $N^r_{n}$
at low $r$ and $n$.
These are in agreement with the known result
(see \cite{Katz:1999xq,Aganagic:2002qg}, for example)
of the Gopakumar--Vafa
invariants for local $\bbP^2$.

\begin{table}[t]
\[
\begin{array}{|c|crrrrrr|}
\hline
N^r_n&n&1&2&3&4&5&\cdots\\ \hline
r&&&&&&&\\[1ex]
0&& 3 &  - 6 &  27 & - 192 & 1695 &\\ 
1&& 0 & 0 & -10 &  231 & - 4452 &\\
2&& 0 & 0 & 0 & -102 &  5430 &\\
3&& 0 & 0 & 0 & 15 & - 3672 &\\
\vdots&&&&&&&\ddots\\ \hline
\end{array}\nn
\]
\caption{\label{table:P2GVinv}Gopakumar--Vafa
invariants for the local $\bbP^2$.}
\end{table}
%

\subsection{Local $\bbP^1\times\bbP^1$}

The coefficients of the Seiberg--Witten curve are given by
\eqb
f\Eqn{=}\frac{1}{12}u^4-\frac{2}{3}\chi u^2
  +\frac{4}{3}\chi^2-4,\nn\\
g\Eqn{=}\frac{1}{216}u^6-\frac{1}{18}\chi u^4
  +\left(\frac{2}{9}\chi^2-\frac{1}{3}\right)u^2
  +\left(-\frac{8}{27}\chi^3+\frac{4}{3}\chi\right),
\eqe
where
\[
\chi=e^{2\pi i \mu}+e^{-2\pi i \mu}.
\]
The amplitudes at $g=0,1$ are given by
(\ref{dPF0}), (\ref{dPF1}) with $n=1$.
Substituting the above $f,\,g$ into (\ref{F2infg}), one obtains
\eqb
\hcalF_2\Eqn{=}
 \frac{1}{12960(u^2-4\chi+8)^2(u^2-4\chi-8)^2}
 \Bigl(100u^2\Gen^3\nn\\
&&\hspace{2em}
  +120\left(-2u^4+5\chi u^2+12\chi^2-48\right)\Gen^2\nn\\
&&\hspace{2em}
  +15\left(13u^6-80\chi u^4+(16\chi^2+768)u^2
           +384\chi^3-1536\chi\right)\Gen\nn\\
&&\hspace{2em}
  +8\left(10u^8-201u^6\chi+(1452\chi^2-2808)u^4\right.\nn\\
&&\hspace{5em}\left.
    +(-4528\chi^3+20304\chi)u^2
    +5184\chi^4-36288\chi^2+62208
   \right)
\Bigr).\qquad
\eqe
We do not present the explicit form of $\hcalF_3$
since it is slightly lengthy,
but the calculation is straightforward.
The instanton expansions in this case read
\eqb
\frac{1}{u}\Eqn{=}
 e^{\phi} - \chi e^{3\phi} +(\chi^2-3) e^{5\phi}
 +(-\chi^3+\chi) e^{7\phi}
 +{\cal O}\left(e^{9\phi}\right),\\
\omega\Eqn{=}
 e^{\phi} + \chi e^{3\phi} + (\chi^2+9) e^{5\phi}
 +(\chi^3+43\chi) e^{7\phi}
 +{\cal O}\left(e^{9\phi}\right),\\
t\Eqn{=}8\phi
 +8\chi e^{2\phi} + (4\chi^2+56) e^{4\phi}
 + \left(\frac{8}{3}\chi^3+208\chi\right)e^{6\phi}
 +{\cal O}\left(e^{8\phi}\right).
\eqe
The all-genus topological string partition function
can be expressed as
\[
\sum_{g=0}^\infty \calF_g x^{2g-2}
=
 \sum_{r=0}^\infty\sum_{n_1,n_2=0}^\infty
 N^r_{n_1,n_2}
 \sum_{m=1}^\infty\frac{1}{m}
 \left(2\sin\frac{mx}{2}\right)^{2r-2}
 Q_1^{mn_1}Q_2^{mn_2},
\]
where
\[
Q_1=e^{2\pi i(2\varphi+\mu)},\qquad
Q_2=e^{2\pi i(2\varphi-\mu)}.
\]
We checked that
$N^r_{n_1,n_2}$ are in agreement with
the known data of the Gopakumar--Vafa invariants
for the local $\bbP^1\times \bbP^1$
(see \cite{Aganagic:2002qg}, for example).

\section{Conclusion and discussion}

In this paper we have developed
a general method of computing topological string amplitudes
for the local $\frac{1}{2}$K3 surface.
We have demonstrated that the amplitudes can be concisely expressed 
in terms of the Seiberg--Witten curve,
which manifestly exhibits good modular properties
and the affine $E_8$ Weyl group invariance.
We have clarified the general structure of the amplitudes.
The amplitudes at $g=0,1$ are given in
(\ref{F0sol}), (\ref{F1sol}),
while higher-genus amplitudes $\hcalF_g\ (g\ge 2)$
are written as a polynomial in
generators expressed in terms of the Seiberg--Witten curve.
Given the structure, one can determine the coefficients
of the polynomials by solving the holomorphic anomaly equation
and the gap condition.
We have explicitly computed the form of the amplitudes
for $g=2,3$.
We have also found that the holomorphic anomaly equation
takes a very simple form if we adopt notations
in which the amplitudes at low genus are slightly modified.

The topological strings on the local $\frac{1}{2}$K3 surface
encompass those on all local del Pezzo surfaces.
We have elucidated
how to reduce the amplitudes
to those for the local del Pezzo surfaces.
By way of illustration,
we have explicitly constructed
the amplitudes for three simple cases.
These amplitudes correctly reproduce
the known Gopakumar--Vafa invariants.

There are several directions for further investigation.
We have proposed that the conjectures
(\ref{Fgstructureinot}), (\ref{Fgstructureinfg})
on the structure of the amplitudes
hold for general $g$.
It is important to prove them
and clarify how they are related to the general scheme
of the polynomial structure \cite{Alim:2007qj}.
Another point to be clarified is the
precise conditions
needed to determine the amplitudes at arbitrarily high genus.
For general $g$, the gap condition used in this paper
is not likely to be sufficient for fixing the amplitude.
On the other hand, it is known that 
regularity at the orbifold point and the large radius point
and the leading behavior at the conifold points
suffice to determine the holomorphic ambiguities
at least for local del Pezzo surfaces
with one or two moduli parameters \cite{Haghighat:2008gw}.
We expect that the same sort of argument
will apply to the case of
the most general local $\frac{1}{2}$K3 surface.

The direct integration method has been applied to
the four-dimensional $\grp{SU}(2)$ 
Seiberg--Witten theories with matters
\cite{Huang:2006si,Huang:2009md,Huang:2011qx}.
We know from Nekrasov partition functions that 
by taking a certain limit
topological string amplitudes
on the toric del Pezzo surfaces reproduce
the prepotential and the gravitational corrections
of the four-dimensional theories.
It is interesting to see how our general formulas
reproduce those results.
The cases of non-toric local del Pezzo surfaces
are of particular interest.
In terms of the Seiberg--Witten curves,
we know how the four-dimensional $\grp{SU}(2)$
theories with an $E_n$ global symmetry
\cite{Minahan:1996fg,Minahan:1996cj,Noguchi:1999xq}
are reproduced from the five-dimensional ones
\cite{Minahan:1997ch,Eguchi:2002fc}.
It would be interesting to construct
the gravitational corrections to
these four-dimensional theories with an $E_n$ flavor symmetry.

The topological recursion \cite{Eynard:2007kz},
or more specifically
the ``remodeling the B-model'' conjecture \cite{Bouchard:2007ys},
is a powerful method of computing topological string amplitudes.
This method is free of the holomorphic ambiguity
and also computes the open string amplitudes.
It would be very interesting
if our expressions for the amplitudes $\hcalF_g$
can be derived by a method
similar to the topological recursion.

\vspace{3ex}

\begin{center}
  {\bf Acknowledgments}
\end{center}

The author was the Yukawa Fellow and this work was supported
in part by the Yukawa Memorial Foundation.
This work was also
supported in part by a Grant-in-Aid
for Scientific Research from the Japan Ministry of Education, Culture, 
Sports, Science and Technology.

\vspace{3ex}

\newpage
\appendix

\section{Seiberg--Witten curve for E-string theory}

The low-energy effective theory of the E-string theory
in $\bbR^4\times T^2$ is described as $\grp{SU}(2)$
Seiberg--Witten theory with nine parameters,
$\tau$ and $\vecm=(\mu_1,\ldots,\mu_8)$.
$\tau$ is regarded as the bare gauge coupling
and $\vecm$ are the masses of fundamental matters.
The theory possesses an $E_8$ flavor symmetry,
and the Weyl group $\WeylEei$ acts on $\vecm$
as an automorphism.
On the other hand,
from the point of view of the six-dimensional theory,
$\tau$ is the modulus of the $T^2$ in the 5,6-directions
and the $\vecm$ are interpreted as Wilson lines along these
directions. The theory therefore admits modular properties
in $\tau$ and double periodicity in $\vecm$.
These symmetries become manifest if we express
the dependence on these parameters through
$\WeylEei$-invariant Jacobi forms.

\subsection{$\WeylEei$-invariant Jacobi forms}

Let $\varphi_{k,m}(\tau,\vecm)$ denote
$\WeylEei$-invariant Jacobi forms of weight $k$ and index $m$.
They are holomorphic in
$\tau\, ({\rm Im}\,\tau>0)$, $\vecm\in\bbC^8$,
and satisfy
the following properties \cite{EichlerZagier, Wirthmuller}:
\renewcommand{\theenumi}{\roman{enumi}}
\renewcommand{\labelenumi}{\theenumi)}
\begin{enumerate}
\item Weyl invariance:
\[
\varphi_{k,m}(\tau,w(\vecm)) = \varphi_{k,m}(\tau,\vecm),\qquad
w\in \WeylEei.
\]

\item Quasi-periodicity:
\[
\varphi_{k,m}(\tau,\vecm+\bvec{v}+\tau\bvec{w})
=e^{-m \pi i (\tau\bvec{w}^2+2\vecm\cdot\bvec{w})}
\varphi_{k,m}(\tau,\vecm),\qquad \bvec{v},\bvec{w}\in \Gamma_8.
\]

\item Modular properties:
\[
\varphi_{k,m}\left(
\frac{a\tau+b}{c\tau+d}\,,\frac{\vecm}{c\tau+d}\right)
=(c\tau+d)^k\exp\left(m\pi i\frac{c}{c\tau+d}\,\vecm^2\right)
\varphi_{k,m}(\tau,\vecm).
\]

\item $\varphi_{k,m}(\tau,\vecm)$ admit a Fourier expansion as
\[
\varphi_{k,m}(\tau,\vecm)
=\sum_{l=0}^\infty
 \sum_{\shortstack[c]{\scriptsize
       $\bvec{v}\in \Gamma_8$\\ \scriptsize$\bvec{v}^2\le 2ml$}}
 c(l,\bvec{v})e^{2\pi i(l\tau+\bvec{v}\cdot\vecm)}.
\]
\end{enumerate}
Here,
$\Gamma_8$ is the $E_8$ root lattice
and 
$
\Bigl(\begin{array}{cc}a&b\\ c&d\end{array}\Bigr)
\in \grp{SL}(2,\bbZ).
$
Note that in this convention the index $m$ coincides with the level
of the affine $E_8$ Lie algebra.

Among others, 
the most fundamental $\WeylEei$-invariant Jacobi form is 
the theta function associated with the lattice $\Gamma_8$,
\[
\ETheta(\tau,\vecm)
=\sum_{\vecw\in\Gamma^8}
   \exp\left(\pi i\tau\vecw^2
   +2\pi i\vecm\cdot\vecw\right)
=\frac{1}{2}\sum_{k=1}^4\prod_{j=1}^8\varth_k(\mu_j,\tau).
\]
One can see 
from the properties of the Jacobi theta functions
that
$\ETheta(\tau,\vecm)$ is
of weight 4 and index 1.
Jacobi forms of higher indices
can be constructed from $\ETheta(\tau,\vecm)$
as follows.

To construct more general $\WeylEei$-invariant Jacobi forms,
we introduce the functions
\eqb
e_1(\tau)\Eqn{=}
 \tfrac{1}{12}\left(\varth_3(\tau)^4+\varth_4(\tau)^4\right),\nn\\
e_2(\tau)\Eqn{=}
 \tfrac{1}{12}\left(\varth_2(\tau)^4-\varth_4(\tau)^4\right),\nn\\
e_3(\tau)\Eqn{=}
 \tfrac{1}{12}\left(-\varth_2(\tau)^4-\varth_3(\tau)^4\right),
\eqe
and
\[
h(\tau)
  =\varth_3(2\tau)\varth_3(6\tau)+\varth_2(2\tau)\varth_2(6\tau).
\]
Let us then define the following
nine $\WeylEei$-invariant Jacobi forms:
\eqb
\vbox{
\baselineskip=4.5ex
\halign{\hspace{-.5em}#&&$\hfil#$&${}#{}$&$#\quad\,\hfil$\cr
&\phantom{B_2(\tau,\vecm)}&&&&&&&&\cr\noalign{\vskip-4.5ex}
&A_1(\tau,\vecm)&=&\ETheta(\tau,\vecm),&
 A_2(\tau,\vecm)&=&\tfrac{8}{9}\Hecke{\ETheta(2\tau,2\vecm)},
  \hspace{-.5em}&
 A_3(\tau,\vecm)&=&\tfrac{27}{28}\Hecke{\ETheta(3\tau,3\vecm)},
  \hspace{-2.2em}\cr
&A_4(\tau,\vecm)&=&\ETheta(\tau,2\vecm),\hspace{-.5em}&
 A_5(\tau,\vecm)&=&\tfrac{125}{126}\Hecke{\ETheta(5\tau,5\vecm)},
  \hspace{-.5em}&&\cr}
\halign{\hspace{-.5em}#&&$\hfil#$&${}#{}$&$#\quad\,\hfil$\cr
&B_2(\tau,\vecm)&=&
  \tfrac{32}{5}\Hecke{e_1(\tau)\ETheta(2\tau,2\vecm)},&
 B_3(\tau,\vecm)&=&
  \tfrac{81}{80}\Hecke{h(\tau)^2\ETheta(3\tau,3\vecm)},\cr
&B_4(\tau,\vecm)&=&
  \tfrac{16}{15}\Hecke{\varth_4(2\tau)^4\ETheta(4\tau,4\vecm)},&
 B_6(\tau,\vecm)&=&
  \tfrac{9}{10}\Hecke{h(\tau)^2\ETheta(6\tau,6\vecm)}.\cr}}
\hspace{-1.2em}
\eqe
Here, $\Hecke{\cdot}$ denotes
the sum of all possible
distinct $\grp{SL}(2,\bbZ)$ transforms of the argument.
Explicitly, they read
\eqb
A_1(\tau,\vecm)\Eqn{=}\ETheta(\tau,\vecm),\qquad
A_4(\tau,\vecm)=\ETheta(\tau,2\vecm),\nn\\
A_n(\tau,\vecm)\Eqn{=}\tfrac{n^3}{n^3+1}\left(
  \ETheta(n\tau,n\vecm)
  +\tfrac{1}{n^4}\mbox{$\sum_{k=0}^{n-1}$}
  \ETheta(\tfrac{\tau+k}{n},\vecm)
 \right),\qquad n=2,3,5,\nn\\
B_2(\tau,\vecm)\Eqn{=}\tfrac{32}{5}\left(
 e_1(\tau)\ETheta(2\tau,2\vecm)
 +\tfrac{1}{2^4}e_3(\tau)\ETheta(\tfrac{\tau}{2},\vecm)
 +\tfrac{1}{2^4}e_2(\tau)\ETheta(\tfrac{\tau+1}{2},\vecm)\right),\nn\\
B_3(\tau,\vecm)\Eqn{=}\tfrac{81}{80}\left(
 h(\tau)^2\ETheta(3\tau,3\vecm)
  -\tfrac{1}{3^5}\mbox{$\sum_{k=0}^{2}$}h(\tfrac{\tau+k}{3})^2
  \ETheta(\tfrac{\tau+k}{3},\vecm)\right),\nn\\
B_4(\tau,\vecm)\Eqn{=}\tfrac{16}{15}\left(
 \varth_4(2\tau)^4\ETheta(4\tau,4\vecm)
 -\tfrac{1}{2^4}\varth_4(2\tau)^4
  \ETheta(\tau+\tfrac{1}{2},2\vecm)\right.\nn\\
&&\hspace{2em}
 \left.
 -\tfrac{1}{2^2\cdot 4^4}\mbox{$\sum_{k=0}^{3}$}
  \varth_2(\tfrac{\tau+k}{2})^4
  \ETheta(\tfrac{\tau+k}{4},\vecm)\right),\nn\\
B_6(\tau,\vecm)\Eqn{=}\tfrac{9}{10}\left(
  h(\tau)^2\ETheta(6\tau,6\vecm)
 +\tfrac{1}{2^4}\mbox{$\sum_{k=0}^{1}$}
  h(\tau+k)^2\ETheta(\tfrac{3\tau+3k}{2},3\vecm)\right.\nn\\
&&\hspace{2em}\left.
 -\tfrac{1}{3\cdot 3^4}\mbox{$\sum_{k=0}^{2}$}
  h(\tfrac{\tau+k}{3})^2\ETheta(\tfrac{2\tau+2k}{3},2\vecm)\right.\nn\\
&&\hspace{2em}\left.
 -\tfrac{1}{3\cdot 6^4}\mbox{$\sum_{k=0}^{5}$}
  h(\tfrac{\tau+k}{3})^2\ETheta(\tfrac{\tau+k}{6},\vecm)\right).
\eqe
$A_n,B_n$ are of index $n$ and weight $4,6$, respectively.
If we set $\vecm=\veczero$,
these Jacobi forms reduce to ordinary modular forms.
We have determined the normalization of $A_n,B_n$
so that they reduce to the Eisenstein series
\[
A_n(\tau,\veczero)=E_4(\tau),\qquad
B_n(\tau,\veczero)=E_6(\tau).
\]

$A_n,B_n$ generate all the $\WeylEei$-invariant Jacobi forms
appearing in the coefficients
of the Seiberg--Witten curve.\footnote{
There are alternative choices for the generators $A_n,B_n$.
For instance,
one can take
$\tfrac{256}{45}\Hecke{e_1(\tau)\ETheta(4\tau,4\vecm)}$
instead of $B_4$ and/or
$\tfrac{54}{55}\Hecke{h(2\tau)^2\ETheta(6\tau,6\vecm)}$
instead of $B_6$.}

\subsection{Seiberg--Witten curve}

The Seiberg--Witten curve for the E-string theory
was constructed in \cite{Eguchi:2002fc}.
Here we present the same curve
expressed in terms of the $\WeylEei$-invariant Jacobi forms
introduced above:
\[
y^2=4x^3-fx-g,
\]
\[
f=\sum_{j=0}^4 a_j u^{4-j},\qquad
g=\sum_{j=0}^6 b_j u^{6-j},
\]
\eqb
a_0\Eqn{=}\frac{1}{12}A_0,\qquad
a_1=0,\qquad
a_2=\frac{6}{E_4\Delta}\Bigl(-A_0A_2+A_1^2\Bigr),\nn\\
a_3\Eqn{=}\frac{1}{9E_4^2\Delta^2}
  \Bigl(-7A_0^2B_0A_3-20A_0^3B_3
       -9A_0B_0A_1A_2+30A_0^2A_1B_2+6B_0A_1^3\Bigr),\nn\\
a_4\Eqn{=}\frac{1}{864E_4^3\Delta^3}
  \Bigl((A_0^6-A_0^3B_0^2)A_4+(56A_0^5-56A_0^2B_0^2)A_1A_3
  -27A_0^5A_2^2\nn\\
&&\hspace{1.5em}
  -90A_0^3B_0A_2B_2-75A_0^4B_2^2+(180A_0^4-36A_0B_0^2)A_1^2A_2\nn\\
&&\hspace{1.5em}
  +240A_0^2B_0A_1^2B_2
  +(-210A_0^3+18B_0^2)A_1^4\Bigr),\nn\\
b_0\Eqn{=}\frac{1}{216}B_0,\qquad
b_1=-\frac{4}{E_4}A_1,\qquad
b_2=\frac{5}{6E_4^2\Delta}\Bigl(A_0^2B_2-B_0A_1^2\Bigr),\nn\\
b_3\Eqn{=}\frac{1}{108E_4^3\Delta^2}
  \Bigl(-7A_0^5A_3-20A_0^3B_0B_3\nn\\
&&\hspace{1.5em}
  -9A_0^4A_1A_2+30A_0^2B_0A_1B_2+(16A_0^3-10B_0^2)A_1^3\Bigr),\nn\\
b_4\Eqn{=}\frac{1}{1728E_4^4\Delta^3}
  \Bigl((-5A_0^7+5A_0^4B_0^2)B_4
  +(80A_0^6-80A_0^3B_0^2)A_1B_3\nn\\
&&\hspace{1.5em}
  +9A_0^5B_0A_2^2+30A_0^6A_2B_2+25A_0^4B_0B_2^2
  -48B_0A_0^4A_1^2A_2\nn\\
&&\hspace{1.5em}
  +(-140A_0^5+60A_0^2B_0^2)A_1^2B_2
  +(74A_0^3B_0-10B_0^3)A_1^4\Bigr),\nn\\
b_5\Eqn{=}\frac{1}{72E_4^5\Delta^3}\Bigl(
  (-21A_0^7+21A_0^4B_0^2)A_5-294A_0^6A_2A_3-770A_0^4B_0B_2A_3\nn\\
&&\hspace{1.5em}
  -840A_0^4B_0A_2B_3-2200A_0^5B_2B_3+168A_0^5A_1^2A_3
  +480B_0A_0^3A_1^2B_3\nn\\
&&\hspace{1.5em}
  -621A_0^5A_1A_2^2+3525A_0^4A_1B_2^2
  +1224A_0^4A_1^3A_2-240A_0^2B_0A_1^3B_2\nn\\
&&\hspace{1.5em}
  +(-456A_0^3+24B_0^2)A_1^5
\Bigr),\nn\\
\noalign{\break}
b_6\Eqn{=}\frac{1}{13436928E_4^6\Delta^5}\Bigl(
  (-20A_0^{12}+40A_0^9B_0^2-20A_0^6B_0^4)B_6\nn\\
&&\hspace{1.5em}
 +(-189A_0^{10}B_0+378A_0^7B_0^3-189A_0^4B_0^5)A_1A_5\nn\\
&&\hspace{1.5em}
 +(-9A_0^{10}B_0+9A_0^7B_0^3)A_2A_4
 +(-15A_0^{11}+15A_0^8B_0^2)B_2A_4\nn\\
&&\hspace{1.5em}
 +(-180A_0^{11}+180A_0^8B_0^2)A_2B_4
 +(-300A_0^9B_0+300A_0^6B_0^3)B_2B_4\nn\\
&&\hspace{1.5em}
 +(22A_0^9B_0-22A_0^6B_0^3)A_1^2A_4
 +(150A_0^{10}+120A_0^7B_0^2-270A_0^4B_0^4)A_1^2B_4\nn\\
&&\hspace{1.5em}
 +(196A_0^{10}B_0-196A_0^7B_0^3)A_3^2
 +(1120A_0^{11}-1120A_0^8B_0^2)A_3B_3\nn\\
&&\hspace{1.5em}
 +(1600A_0^9B_0-1600A_0^6B_0^3)B_3^2
 +(-2982A_0^9B_0+2982A_0^6B_0^3)A_1A_2A_3\nn\\
&&\hspace{1.5em}
 +(-2520A_0^{10}-4410A_0^7B_0^2+6930A_0^4B_0^4)A_1B_2A_3\nn\\
&&\hspace{1.5em}
 +(3360A_0^{10}-10920A_0^7B_0^2+7560A_0^4B_0^4)A_1A_2B_3\nn\\
&&\hspace{1.5em}
 +(-19800A_0^8B_0+19800A_0^5B_0^3)A_1B_2B_3
 +(2016A_0^8B_0-2016A_0^5B_0^3)A_1^3A_3\nn\\
&&\hspace{1.5em}
 +(-5920A_0^9+7360A_0^6B_0^2-1440A_0^3B_0^4)A_1^3B_3
 +(405A_0^9B_0+162A_0^6B_0^3)A_2^3\nn\\
&&\hspace{1.5em}
 +(1215A_0^{10}+1620A_0^7B_0^2)A_2^2B_2
 +4725A_0^8B_0A_2B_2^2\nn\\
&&\hspace{1.5em}
 +(1125A_0^9+1500A_0^6B_0^2)B_2^3
 +(-9477A_0^8B_0+5103A_0^5B_0^3)A_1^2A_2^2\nn\\
&&\hspace{1.5em}
 +(-9180A_0^9-5400A_0^6B_0^2)A_1^2A_2B_2
 +(20925A_0^7B_0-33075A_0^4B_0^3)A_1^2B_2^2\nn\\
&&\hspace{1.5em}
 +(20304A_0^7B_0-9072A_0^4B_0^3)A_1^4A_2\nn\\
&&\hspace{1.5em}
 +(12780A_0^8+5400A_0^5B_0^2+540A_0^2B_0^4)A_1^4B_2\nn\\
&&\hspace{1.5em}
 +(-11076A_0^6B_0+1512A_0^3B_0^3-36B_0^5)A_1^6
\Bigr).
\eqe

Note that
$a_n,b_n$ satisfy most of the properties
of the $\WeylEei$-invariant Jacobi forms
except the condition $\bvec{v}^2\le 2ml$
in the Fourier expansion.
$a_n,b_n$ are
of index $n$ and weight $4-6n,6-6n$, respectively.
It is useful to let the variables $u,x,y$
transform formally as Jacobi forms of
weights $-6,-10,-15$ and index $1,2,3$, respectively.
The whole curve
then transforms as a Jacobi form of weight $-30$ and index $6$.
$f,g$ are of weight $-20,-30$ and index $4,6$, respectively.

\newpage
\section{Derivative formulas}

%
\eqb
q\frac{d}{dq} \ln\Delta\Eqn{=}E_2,\\
q\frac{d}{dq} E_2\Eqn{=}\frac{1}{12}(E_2^2-E_4),\\
q\frac{d}{dq} E_4\Eqn{=}\frac{1}{3}(E_4E_2-E_6),\\
q\frac{d}{dq} E_6\Eqn{=}\frac{1}{2}(E_6E_2-E_4^2).
\eqe
\eqb
\label{diffE2first}
\left(\dbE t\right)_u\Eqn{=}-2t^2,\\
\left(\dbE\omega\right)_u\Eqn{=}2\omega t,\\
\left(\dbE\phi\right)_u\Eqn{=}2\dphi^{-1}t,\\
\left(\dbE\ln\tDelta\right)_u\Eqn{=}24t,\\
\dbE\tE_{2k}\Eqn{=}4kt\tE_{2k}+24\delta_{1,k}\,.
\label{diffE2last}
\eqe
\eqb
\label{diffE2bisfirst}
\left(\dbE\left(\dphi^n\ln\omega\right)\right)_u\Eqn{=}
 -2\sum_{k=0}^{n-1}
  \left(\begin{array}{@{}c@{}}n\\[.5ex] k+1\end{array}\right)
  \dphi^k t\,\dphi^{n-k}\ln\omega
 +2\dphi^n t,\\
\left(\dbE\left(\dphi^n t\right)\right)_u\Eqn{=}
 -2\sum_{k=0}^{n-1}
  \left[
  \left(\begin{array}{@{}c@{}}n\\[.5ex] k+1\end{array}\right)
  +2\left(\begin{array}{@{}c@{}}n-1\\[.5ex] k\end{array}\right)
\right]
  \dphi^k t\,\dphi^{n-k} t\qquad (n\ge 1).\qquad
\label{diffE2bislast}
\eqe
%

\section{Genus three amplitude}

%
\eqb
\calF_3\Eqn{=}
\phantom{+}(\dphi^4\ln\omega)
 \left(
   \tfrac{1}{2304}\tE_2^2
  +\tfrac{1}{2592}\tE_4
 \right)\nn\\
&&
+(\dphi^3\ln\omega)(\dphi\ln\omega)
 \left(
   \tfrac{1}{1152}\tE_2^2
  -\tfrac{1}{6912}\tE_4
 \right)\nn\\
&&
+(\dphi^2\ln\omega)^2
 \left(
   \tfrac{1}{2304}\tE_2^2
  +\tfrac{11}{20736}\tE_4
 \right)\nn\\
&&
+(\dphi^2\ln\omega)(\dphi\ln\omega)^2
 \left(
   \tfrac{1}{2304}\tE_2^2
  +\tfrac{1}{20736}\tE_4
 \right)\nn\\
&&
+(\dphi^3\ln\omega)(\dphi t)
 \left(
   \tfrac{1}{41472}\tE_2^3
  +\tfrac{11}{82944}\tE_4\tE_2
  -\tfrac{13}{82944}\tE_6
 \right)\nn\\
&&
+(\dphi^2\ln\omega)(\dphi\ln\omega)(\dphi t)
 \left(
   \tfrac{1}{13824}\tE_2^3
  -\tfrac{25}{248832}\tE_4\tE_2
  +\tfrac{7}{248832}\tE_6
 \right)\nn\\
&&
+(\dphi\ln\omega)^3(\dphi t)
 \left(
   \tfrac{1}{41472}\tE_2^3
  -\tfrac{1}{31104}\tE_4\tE_2
  +\tfrac{1}{124416}\tE_6
 \right)\nn\\
&&
+(\dphi^2\ln\omega)(\dphi^2t)
 \left(
   \tfrac{7}{62208}\tE_4\tE_2
  -\tfrac{7}{62208}\tE_6
 \right)\nn\\
&&
+(\dphi^2\ln\omega)(\dphi t)^2
 \left(
  -\tfrac{1}{331776}\tE_2^4
  +\tfrac{79}{1492992}\tE_4\tE_2^2
  -\tfrac{35}{373248}\tE_6\tE_2
  +\tfrac{131}{2985984}\tE_4^2
 \right)\nn\\
&&
+(\dphi\ln\omega)^2(\dphi^2t)
 \left(
   \tfrac{5}{248832}\tE_4\tE_2
  -\tfrac{5}{248832}\tE_6
 \right)\nn\\
&&
+(\dphi\ln\omega)^2(\dphi t)^2
 \left(
  -\tfrac{1}{331776}\tE_2^4
  +\tfrac{43}{2985984}\tE_4\tE_2^2
  -\tfrac{25}{1492992}\tE_6\tE_2
  +\tfrac{1}{186624}\tE_4^2
 \right)\nn\\
&&
+(\dphi\ln\omega)(\dphi^3t)
 \left(
  -\tfrac{1}{13824}\tE_2^3
  +\tfrac{1}{19440}\tE_4\tE_2
  +\tfrac{13}{622080}\tE_6
 \right)\nn\\
&&
+(\dphi\ln\omega)(\dphi^2t)(\dphi t)
 \left(
  -\tfrac{5}{165888}\tE_2^4
  +\tfrac{35}{497664}\tE_4\tE_2^2
  -\tfrac{5}{248832}\tE_6\tE_2
  -\tfrac{5}{248832}\tE_4^2
 \right)\nn\\
&&
+(\dphi\ln\omega)(\dphi t)^3
 \left(
  -\tfrac{1}{497664}\tE_2^5
  +\tfrac{29}{2985984}\tE_4\tE_2^3
 \right.\nn\\ &&\hspace{8em}\left.
  -\tfrac{1}{110592}\tE_6\tE_2^2
  -\tfrac{1}{995328}\tE_4^2\tE_2
  +\tfrac{7}{2985984}\tE_6\tE_4
 \right)\nn\\
&&
+(\dphi^4t)
 \left(
  -\tfrac{1}{20736}\tE_2^3
  -\tfrac{121}{1244160}\tE_4\tE_2
  -\tfrac{173}{8709120}\tE_6
 \right)\nn\\
&&
+(\dphi^3t)(\dphi t)
 \left(
  -\tfrac{11}{497664}\tE_2^4
  -\tfrac{287}{3732480}\tE_4\tE_2^2
  +\tfrac{421}{6531840}\tE_6\tE_2
  +\tfrac{361}{10450944}\tE_4^2
 \right)\nn\\
&&
+(\dphi^2t)^2
 \left(
  -\tfrac{1}{55296}\tE_2^4
  -\tfrac{19}{331776}\tE_4\tE_2^2
  +\tfrac{19}{387072}\tE_6\tE_2
  +\tfrac{61}{2322432}\tE_4^2
 \right)\nn\\
&&
+(\dphi^2t)(\dphi t)^2
 \left(
  -\tfrac{13}{1990656}\tE_2^5
  -\tfrac{1}{27648}\tE_4\tE_2^3
\right.\nn\\ &&\hspace{8em}\left.
  +\tfrac{25}{331776}\tE_6\tE_2^2
  -\tfrac{19}{1990656}\tE_4^2\tE_2
  -\tfrac{23}{995328}\tE_6\tE_4
 \right)\nn\\
&&
+(\dphi t)^4
 \left(
  -\tfrac{7}{23887872}\tE_2^6
  -\tfrac{181}{71663616}\tE_4\tE_2^4
  +\tfrac{19}{2239488}\tE_6\tE_2^3
\right.\nn\\ &&\hspace{8em}\left.
  -\tfrac{47}{7962624}\tE_4^2\tE_2^2
  -\tfrac{1}{559872}\tE_6\tE_4\tE_2
  +\tfrac{73}{71663616}\tE_4^3
  +\tfrac{1}{995328}\tE_6^2
 \right). \nn\\
\eqe
%

\section{Conventions}

We define the Eisenstein series, the modular discriminant,
and the $j$-invariant
by their Fourier expansion:
\eqb
E_{2n}(\tau)\Eqn{=}1+\frac{(2\pi i)^{2n}}{(2n-1)!\,\zeta(2n)}
\sum_{k=1}^{\infty}\frac{k^{2n-1}q^k}{1-q^k},
\qquad q=e^{2\pi i\tau},\\[1ex]
\Delta(\tau)\Eqn{=}
 q\left[\tprod_{k=1}^\infty(1-q^k)\right]^{24}
=\frac{1}{1728}\Bigl({E_4(\tau)}^3-{E_6(\tau)}^2\Bigr),\\[1ex]
j(\tau)\Eqn{=}\frac{E_4(\tau)^3}{\Delta(\tau)}.
\eqe
We often omit the argument of these functions, as far as it is $\tau$.
When the argument is $\ttau$, we
use the following abbreviations:
\[
\tE_{2n}:=E_{2n}(\ttau),\qquad
\tDelta:=\Delta(\ttau),\qquad
\tj:=j(\ttau).
\]

The Weierstrass $\wp$-function is defined as
\[ 
\wp(z|2\pi\omega,2\pi\omega\tau)
=\frac{1}{z^2}+\!\!
\sum_{m,n\in\bbZ^2_{\ne (0,0)}}\!\!\left[\frac{1}{(z-\Omega_{m,n})^2}
  -\frac{1}{{\Omega_{m,n}}^2}\right],
\quad \Omega_{m,n}=2\pi\omega(m+n\tau).
\]
This function
satisfies the 
differential equation
\[
(\partial_z \wp)^2=4\wp^3-\frac{E_4(\tau)}{12\omega^4}\wp
               -\frac{E_6(\tau)}{216\omega^6}.
\]

The Jacobi theta functions are defined as
\eqb
\varth_1(z,\tau)\Eqn{=}
  i\sum_{n\in \bbZ}(-1)^n y^{n-1/2}q^{(n-1/2)^2/2},\\
\varth_2(z,\tau)\Eqn{=}
  \sum_{n\in \bbZ}y^{n-1/2}q^{(n-1/2)^2/2},\\
\varth_3(z,\tau)\Eqn{=}
  \sum_{n\in \bbZ}y^n q^{n^2/2},\\
\varth_4(z,\tau)\Eqn{=}
  \sum_{n\in \bbZ}(-1)^n y^n q^{n^2/2},
\eqe
where $y=e^{2\pi i z},\ q=e^{2\pi i \tau}$.
We also use the following abbreviated notation:
\[
\varth_k(\tau):=\varth_k(0,\tau).
\]
%

\newpage

\renewcommand{\section}{\subsection}
\renewcommand{\refname}

\end{document}